\documentclass[aps,prb,twocolumn,showpacs,superscriptaddress]{revtex4}
\usepackage{amssymb,amsmath}
\usepackage{graphicx}
\usepackage{multirow}
\usepackage{hyperref}
\usepackage{appendix}
\usepackage{mathtools}
\usepackage{amsmath}
\usepackage{xcolor}
\usepackage{siunitx}
\graphicspath{{./img/}}

\begin{document}
\title{
Polarons in the Cubic Generalized Fr\"ohlich Model: Spontaneous Symmetry Breaking
}

\author{Vasilii Vasilchenko}
\affiliation{European Theoretical Spectroscopy Facility, Institute of Condensed Matter and Nanosciences, Universit\'{e} catholique de Louvain, Chemin des \'{e}toiles 8, bte L07.03.01, B-1348 Louvain-la-Neuve, Belgium}
\author{Xavier Gonze}
\affiliation{European Theoretical Spectroscopy Facility, Institute of Condensed Matter and Nanosciences, Universit\'{e} catholique de Louvain, Chemin des \'{e}toiles 8, bte L07.03.01, B-1348 Louvain-la-Neuve, Belgium}


\begin{abstract}
Within the variational polaron equation framework, the Fr\"ohlich model for cubic systems with three-fold degenerate electronic bands is numerically solved in the strong coupling regime, for a wide range of its input parameters.
By comparing the results to the previously reported ones obtained with the Gaussian Ansatz approach, the inadequacy of the latter is uncovered, especially when degenerate bands are present in a system.
Moreover, the symmetry groups of polaronic solutions in the cubic generalized Fr\"ohlich model without spin-orbit coupling are investigated: we provide and discuss a phase diagram of symmetry groups of ground-state polarons, showing
 spontaneous symmetry breaking.
While the cubic symmetry of the three-band degenerate model Hamiltonian corresponds to the full octahedral group $O_h$, lowest-energy polarons possess either $D_{4h}$ or $D_{3d}$ point groups.
This phase diagram bears some similarities but differs nevertheless from the one that is obtained by the straight analysis of the band effective masses.
The obtained results will provide a firm ground for further exploration of the generalized Fr\"ohlich model and will likely be applicable beyond the model's inherent approximations.

\end{abstract} 


\maketitle

\section{Introduction}
\label{sec:introduction}

In many condensed matter systems, electron-phonon interaction (EPI) may lead to the formation of a quasiparticle called polaron.
In this phenomenon, a bare charge carrier creates a field of lattice excitations and becomes dressed by this self-induced phonon cloud.
This process generates lattice deformation, alters the carrier's effective mass, and may lead to its autolocalization in the induced deformation potential.
Polarons are observed in various classes of materials,
\cite{alexandrov_polarons_2007, franchini_polarons_2021}
ranging from bulk crystals \cite{hodby_cyclotron_1967, giannini_quantum_2019, ghosh_polarons_2020, wang_band_2022, rene_de_cotret_direct_2022, redondo_real-space_2023}
to 2D semiconductors
\cite{kang_holstein_2018, vasilchenko_small_2021, liu_atomic-scale_2023, sio_polarons_2023}
by means of both experimental and theoretical methods.
While state-of-the-art experimental techniques allow probing polaronic properties in many systems of interest,
\cite{rene_de_cotret_direct_2022, liu_atomic-scale_2023, redondo_real-space_2023}
\textit{ab initio} methods are rapidly developing
\cite{sio_ab_2019, sio_polarons_2019, lee_facile_2021, vasilchenko_variational_2022, lafuente-bartolome_ab_2022, lafuente-bartolome_unified_2022,  falletta_polarons_2022, sio_polarons_2023}
and paving the way for further advances in the field. 

In most instances, the study of polarons primarily focuses on the interaction between excess electrons or holes and lattice vibrations, giving rise to the notion of electron and hole polarons.
However, the fundamental idea of a particle coupled with phonons holds true regardless of particle statistics, leading to the existence of other types of polarons, such as exciton polarons.
\cite{campbell_exciton-polarons_2022, srimath_kandada_exciton_2020, Dai2024, Dai2024a}
Polarons are commonly characterized as either small or large based on the spatial extent of the associated phonon cloud and associated electronic localization compared to the characteristic interatomic distance.
This classification reflects the distinct nature and physical characteristics of these quasiparticles, which result from variations in the effective range of EPI.
Furthermore, the strength of EPI introduces an additional degree of freedom into the problem, and two limiting regimes can be distinguished: weak and strong coupling.
In the former regime, a carrier coherently drags the accompanying phonon cloud as it moves, whereas, in the latter, it becomes self-trapped within the potential well induced by lattice deformation.
The transitional region between these two scenarios is often referred to as the intermediate coupling regime, which is particularly relevant to many materials and most challenging to address.

Historically, the concept of electron autolocalization in a crystal originates from the work of Landau published in 1933.
\cite{landau_electron_1933}
Subsequently, the term ``polaron'' was introduced by Pekar, who characterized this quasiparticle in the strong-coupling regime by classical treatment of the lattice deformations.
\cite{pekar_local_1946}
The polaron theory further underwent rapid development, leading to the formulation of two well-known quantum-field Hamiltonians by Fr\"ohlich
\cite{frohlich_xx_1950, frohlich_electrons_1954}
and Holstein,
\cite{holstein_studies_1959, holstein_studies_1959-1}
which were designed for large and small polarons, respectively.
Both models have continued to attract substantial attention,
\cite{perroni_spectral_2011, Zhugayevych2015, miglio_predominance_2020, guster_frohlich_2021, vasilchenko_variational_2022, brousseau-couture_effect_2023}
serving as a robust foundation for the all-coupling theory of polarons.
\cite{feynman_slow_1955, mishchenko_diagrammatic_2000, mishchenko_polaron_2019, lafuente-bartolome_ab_2022, lafuente-bartolome_unified_2022}

While these theoretical approaches describe idealized systems, real materials are far more complex.
In practice, taking into account the full complexity of real materials can be achieved through first-principles calculations, such as density functional theory (DFT).
\cite{sio_ab_2019, sio_polarons_2019, lee_facile_2021, vasilchenko_variational_2022, lafuente-bartolome_ab_2022, lafuente-bartolome_unified_2022,  falletta_polarons_2022, sio_polarons_2023}
Nonetheless, model Hamiltonians, despite their simplicity, remain valuable tools for capturing the fundamental physics of polaron formation.
They are particularly useful for establishing benchmarks for \textit{ab initio} formalisms.
Besides, these models can be extended by lifting some of their initial approximations.
This extension brings them closer to explaining the essential electron-phonon effects in real materials while still maintaining their clarity and simplicity.
For example, the Fr\"ohlich model has been extensively investigated since its first formulation in 1950. Still, a recent generalization by Miglio \textit{et al.}
\cite{miglio_predominance_2020}
has paved the way for a fresh venue of research, as described now.
\cite{guster_frohlich_2021, martin_multiple_2023, guster_large_2023, de_melo_high-throughput_2023, brousseau-couture_effect_2023}

The standard Fr\"ohlich formalism assumes the continuum approximation, where an excess charge carrier interacts with an oscillating dielectric continuum rather than the crystal lattice.
This approximation holds true for large polarons.
Within this formalism, it considers the coupling of one dispersionless longitudinal optical (LO) phonon mode to a single electronic band with isotropic dispersion through a screened Couloumb-like EPI.
For both weak and strong coupling scenarios in polaron formation, the model provides well-established asymptotic solutions.
\cite{Roseler1968, j_miyake_ground_1976}
The intermediate regime presents however a more challenging problem and requires non-perturbative methods such as the Feynman path integral approach
\cite{feynman_slow_1955}
or diagrammatic Monte Carlo technique. 
\cite{mishchenko_diagrammatic_2000}

The generalized Fr\"ohlich model of Miglio \textit{et al.}
\cite{miglio_predominance_2020} expands upon the original formalism by considering several degenerate and anisotropic electronic bands, and multiple LO modes.
This broader framework provides a comprehensive description of polaron formation in a variety of polar materials, including cubic ones, and is applicable, for instance, in high-throughput calculations.
\cite{de_melo_high-throughput_2023}
In the weak-coupling regime, the lowest-order perturbation has been employed to study this model possessing an underlying cubic symmetry, based on the three-band Luttinger-Kohn (LK) Hamiltonian
\cite{miglio_predominance_2020, guster_frohlich_2021}
For the strong coupling limit, the initial approach involved variational methods employing the Gaussian Ansatz technique.
\cite{guster_frohlich_2021}
Subsequent investigations have revealed minor deviations from the Gaussian wavefunctions when a fully variational treatment was applied for non-degenerate bands exhibiting moderate anisotropy. 
\cite{vasilchenko_variational_2022}
However, for the case of degenerate bands, the only approach employed has been the use of Gaussian Ansatz for cubic systems, and no fully variational solution has been reported thus far. 

This paper presents a continuation of the ongoing research concerning the variational treatment of the generalized Fr\"ohlich model.
In this work, we consider both the degeneracy and anisotropy of electronic bands in cubic systems and we approach the model using a fully variational formalism,
\cite{vasilchenko_variational_2022}
yielding accurate numerical solutions.
In contrast to the straightforward non-degenerate anisotropic scenario, our findings significantly diverge from Gaussian results.
This highlights the inadequacy of the Gaussian trial wavefunctions for materials featuring degenerate bands. 
Additionally, we observe spontaneous symmetry breaking in the ground-state polaron wavefunction.
Depending on the model parameters, the lowest-energy solution does not share the same symmetry as the one of the generalized Fr\"ohlich Hamiltonian.
This effect is attributed to the band degeneracy, and we provide a corresponding symmetry phase diagram to illustrate this phenomenon.

This spontaneous symmetry breaking belongs to the large set of  Jahn-Teller\cite{Jahn_1937,Bersuker_2006} spontaneous breakings, originating from the electron-vibration coupling when the starting electronic and vibronic ground states are degenerate.
The formation of a polaron in the Fr\"ohlich Hamiltonian spontaneously breaks the separate translational symmetry of the electrons and phonons, hence, it is a manifestation of the  Jahn-Teller effect.
The present situation corresponds to a further symmetry breaking, now related to the point symmetries. 
Models help to understand the underlying physics of this effect, even if not capturing the details of the phenomena in real materials.

In the three-band Luttinger-Kohn Hamiltonian, used in this work, a three-fold degeneracy is present at the zone center.
In real cubic materials, at the top of the valence band, this three-fold degeneracy is lifted due to spin-orbit coupling (SOC). 
One is left with a two-fold degeneracy, for so-called light hole and heavy hole bands, and a spin-orbit split-off non-degenerate band. 
The SOC is properly accounted for by the Dresselhaus model\cite{brousseau-couture_effect_2023}, unlike the Luttinger-Kohn Hamiltonian.
Still, the LK Hamiltonian might be an interesting starting point for materials composed of lighter elements, as the SOC varies strongly with the atomic number. 

Until now, such SOC effects have been considered only within the weak-coupling limit of the generalized Fr\"ohlich model.
\cite{brousseau-couture_effect_2023, Trebin1975}
Our investigation covers the strong-coupling regime, for which prior analyses of the model
\cite{guster_frohlich_2021}
and recent supercell studies of polaron formation uniformly neglected spin-orbit interactions.
\cite{sio_ab_2019, sio_polarons_2019, lee_facile_2021, lafuente-bartolome_ab_2022, lafuente-bartolome_unified_2022,  falletta_polarons_2022, sio_polarons_2023}.
Although relevant for real materials, we will neglect the SOC  in the present study, aiming at establishing the  kind of symmetry breaking that appear in our model, providing a full phase diagram, as well as establishing the aspects in which the variational treatment deviates from the Gaussian results.

The neglect of SOC limits the applicability of the present study for real materials.
For example, the symmetry of the ground-state polaron with and without SOC will likely not be the same. 
Such an analysis is left for further work.

The paper is structured as follows.
In the next section (Sec.~\ref{sec:methodology}) the necessary background information is provided and the notations used throughout the paper are established.
Namely, we first recall the generalized Fr\"ohlich model for cubic systems with degenerate bands and discuss the Luttinger-Kohn Hamiltonian required to initialize the electronic configuration of the model.
This is followed by a recap of the variational approach and a discussion on its application to solve the cubic generalized Fr\"ohlich model in the strong coupling regime.
At the end of the section, a comprehensive analysis of the potential symmetries of the variational solutions is provided.
In Sec.~\ref{sec:setup} we list the technical details concerning the calculations within the variational framework.
Then, in Sec.~\ref{sec:results} the cubic generalized Fr\"ohlich model is variationally solved 
for a wide range of input parameters, and the resulting symmetry of polarons is analyzed.

\section{Methodology}
\label{sec:methodology}

\subsection{Cubic Generalized Fr\"ohlich Model}
\label{sec:gfr-cubic}

The generalized Fr\"ohlich model,\cite{miglio_predominance_2020} introduced by Miglio \textit{et al.}, extends the original Fr\"ohlich formalism while maintaining the underlying continuum hypothesis, which is valid for large polarons.
In this approximation, the system of interest is treated as a dielectric continuum, characterized solely by its macroscopic parameters.

In this work, we focus on the application of the model to systems with cubic symmetry.
These systems are convenient for consideration due to their isotropic dielectric constant and LO phonon modes.
Nonetheless, the key elements of the generalized Fr\"ohlich formalism are retained, including the degeneracy and anisotropy of electronic bands, as well as the coupling to multiple LO phonons.

In what follows, the main features of the cubic generalized Fr\"ohlich model
are outlined.
Throughout the work, the Hartree atomic unit system
is adopted, unless stated otherwise:
$\hbar = m_e = |e| = c = 1$.

The cubic generalized Fr\"ohlich Hamiltonian
\cite{guster_frohlich_2021}
reads as
\begin{equation}
    \hat{H}^\mathrm{gFr}
    = \hat{H}^\mathrm{gFr}_\mathrm{el}
    + \hat{H}^\mathrm{gFr}_\mathrm{ph}
    + \hat{H}^\mathrm{gFr}_\mathrm{el-ph}, 
\end{equation}
with 
\begin{align}
    &
    \hat{H}^\mathrm{gFr}_\mathrm{el}
    = \sum_{n\mathbf{k}}
    \frac{\sigma k^2}{2m^*_n(\mathbf{\hat{k}})}
    \hat{c}^\dagger_{n\mathbf{k}} \hat{c}_{n\mathbf{k}}, \label{gfr:el} \\
    &
    \hat{H}^\mathrm{gFr}_\mathrm{ph}
    = \sum_{\mathbf{q}\nu}
    \omega_{\nu \mathrm{LO}}
    \left(
    \hat{a}^\dagger_{\mathbf{q}\nu} \hat{a}_{\mathbf{q}\nu} + \frac{1}{2}
    \right), \label{gfr:ph} \\
    &
    \hat{H}^\mathrm{gFr}_\mathrm{el-ph}
    = \sum_{\substack{mn\nu \\ \mathbf{k q}}}
    g^\mathrm{gFr}_{mn\nu}(\mathbf{k, q})
    \hat{c}^\dagger_{m\mathbf{k + q}} \hat{c}_{n\mathbf{k}}
    \left(
    \hat{a}_{\mathbf{q}\nu} + \hat{a}^\dagger_{\mathbf{-q}\nu}
    \right), \label{gfr:elph}
\end{align} 
where $\hat{c}^\dagger_{n\mathbf{k}}$/$\hat{c}_{n\mathbf{k}}$ and $\hat{a}^\dagger_{\mathbf{q}\nu}$/$\hat{a}_{\mathbf{q}\nu}$ are electron and phonon creation/annihilation operators, respectively.
Each term is determined based on the parameters initially defined at the zone center $\Gamma$ and subsequently extended to encompass the whole reciprocal space.

Eq.~(\ref{gfr:el}), which is the kinetic energy term, includes parabolic bare electronic energies, characterized by direction-dependent effective masses $m^*_n(\mathbf{\hat{k}})$.
These effective masses, associated with the band index $n$ and wavevector $\mathbf{k}$, are governed by Luttinger-Kohn parameters in case of three-fold degeneracy.
\cite{Luttinger1955, LaflammeJanssen2016}
The parameter $\sigma$ is introduced to ensure positive effective masses near the band extrema and  to characterize the type of electronic bands:
$\sigma = 1$ for conduction bands and $\sigma = -1$ for valence bands.
This distinction corresponds to the electron and hole formation, respectively.

Eq.~(\ref{gfr:ph}) represents the vibrational term, accommodating multiple LO phonon modes of index $\nu$ and wavevector $\mathbf{q}$.
These modes are characterized by dispersionless, direction-independent frequencies $\omega_{{\nu},\mathrm{LO}}$.

The electron-phonon coupling term defined by Eq.~(\ref{gfr:elph}) is determined by the generalized Fr\"ohlich EPI matrix elements.
\cite{Verdi2015}
They describe the scattering process from an electronic state with the band index $n$ and wavevector $\mathbf{k}$ to a state with the band index $n'$ and wavevector $\mathbf{k'} = \mathbf{k + q}$ through a LO phonon of mode $\nu$:
\begin{multline}\label{gfr:epi}
    g_{n'n\nu}^\mathrm{gFr}(\mathbf{k, q})
     = \frac{1}{q} \frac{4\pi}{\Omega_0}
      \left(
      \frac{1}{2\omega_{\nu,\mathrm{LO}}
      N_p }
      \right)^{1/2}
      \frac{p_{\nu, \mathrm{LO}}}{\epsilon^\infty}
      \\
      \times \sum_m
      s_{n'm}(\hat{\mathbf{k'}}) s^*_{nm}(\hat{\mathbf{k}}).
\end{multline}
Here $\Omega_0$ and $N_p$ represent the primitive unit cell volume and the Born-von Karman supercell size, defined by the Brillouin zone (BZ) sampling.
The macroscopic dielectric constant $\epsilon^\infty$ and LO  mode polarities $p_{\nu, \mathrm{LO}}$ are isotropic since cubic systems are considered.
The connection between electronic states is captured by the overlap matrices
\begin{equation}\label{gfr:smatrix}
    s_{nm}(\hat{\mathbf{k}})
    =
    \langle
    n, \hat{\mathbf{k}} |
    m, c
    \rangle_{\mathrm{P}}.
\end{equation}
Here, the subscript P indicates that the integration is performed using the periodic components of Bloch wavefunctions.

In the  Hamiltonian presented above, the summation over electronic bands is limited to the degenerate states linked to the band extremum.
For the sake of convenience, the band extremum is consistently assumed to be located at the $\Gamma$ point,  although it can be altered by modifying the definition of the $s$-matrices in Eq.~(\ref{gfr:smatrix}).
Also, only LO phonons are considered, as all the others are automatically excluded due to the condition $p_\nu = 0$ that holds for non-LO phonons.

The cubic generalized Fr\"ohlich Hamiltonian incorporates (i) multiple LO phonon modes $\omega_{\nu, \mathrm{LO}}$ and (ii) several anisotropic degenerate electronic bands with corresponding effective masses $m^*_n(\mathbf{\hat{k}})$.
This extension represents a significant improvement over the original Fr\"ohlich model, which accounts for only a single LO phonon branch $\omega_\mathrm{LO}$ coupled to a single isotropic band of mass $m^*$. 
Hence, this brings Fr\"ohlich formalism closer to the accurate representation of polarons in real materials.
Notably, the presence of three-fold degeneracy in multiple cubic oxides, II-VI and III-V semiconductors, among others, is a common occurrence.
\cite{miglio_predominance_2020, guster_frohlich_2021}

The model relies on several macroscopic parameters, defining the ground-state electronic and vibrational configuration of a system, along with electron-phonon coupling.
These parameters encompass effective masses $m^*_n(\mathbf{\hat{k}})$, LO phonon modes frequencies $\omega_{\nu, \mathrm{LO}}$ and polarities $p_{\nu, \mathrm{LO}}$, optical dielectric constant $\epsilon^\infty$ and overlap $s$-matrices.
These values are easily obtainable through either first-principles calculations or experimental measurements.

\subsection{Three-fold Degenerate Bands and Luttinger-Kohn Hamiltonian}
\label{sec:lk}

Notably, numerous cubic materials exhibit three-fold degenerate bands at the top of valence bands, and some at the bottom of the conduction bands.
To account for effective masses and overlaps between three-fold degenerate states near band extrema, the Luttinger-Kohn (LK) Hamiltonian can be utilized.
\cite{Luttinger1955}
Given the importance of these parameters to the generalized Fr\"ohlich model outlined above, the subsequent discussion focuses on the key aspects of the LK Hamiltonian and its relevance to the model. 

Derived from the $\mathbf{k}\cdot\mathbf{p}$ perturbation theory, the LK Hamiltonian characterizes the behavior of multiple degenerate electronic bands near band extrema.
For cubic materials exhibiting three-fold band degeneracy at $\Gamma$ the Hamiltonian is given by:
\begin{multline}\label{eq:h-lk}
        \hat{H}^\mathrm{LK} (\mathbf{k})
        =
        \\
        \begin{bsmallmatrix}
            a k_x^2 + b \left( k_y^2 + k_z^2 \right) & c k_x k_y & c k_x k_z \\
            c k_x k_y & a k_y^2 + b \left( k_x^2 + k_z^2 \right) & c k_y k_z \\
            c k_x k_z & c k_y k_z & a k_z^2 + b \left( k_x^2 + k_y^2 \right)
        \end{bsmallmatrix},
\end{multline}
and is determined by the effective mass tensor of a system.
Here, $a$, $b$, and $c$ parameters have dimensions of inverse effective mass and are straightforwardly linked to the effective masses $m^*$ along the three principal directions in reciprocal space, $[100]$, $[110]$, and $[111]$:
\begin{equation}\label{eq:lk_params*}
    \begin{split}
    & (m^*_{100})^{-1}
    =
    \begin{cases}
        2 a, \\
        2 b~\mathrm{(twofold)},
    \end{cases} 
    \\
    & (m^*_{110})^{-1}
    =
    \begin{cases}
        a + b + c, \\
        a + b - c, \\
        2 b,
    \end{cases}
    \\
    & (m^*_{111})^{-1}
    =
    \begin{cases}
        \frac{2}{3} \left( a + 2b + 2c \right), \\
        \frac{2}{3} \left( a + 2b - c \right)~\mathrm{(twofold)}.
    \end{cases}
    \end{split}
\end{equation}
These equations also provide limits for the values of LK parameters, ensuring that effective masses are positive and the Hamiltonian is bounded:
\begin{equation}\label{eq:lk-limits}
    \begin{cases}
        a > 0, \\
        b > 0, \\
        c < a + b, \\
        c > -\frac{1}{2}a - b.
    \end{cases}
\end{equation}

Hence, to set the ground-state electronic configuration of a triply degenerate cubic crystal, one can either rely on the effective masses along the three symmetry-inequivalent directions or, alternatively, employ the $a$, $b$, and $c$ parameters of the LK Hamiltonian.
Diagonalization of the Hamiltonian yields three electronic bands with degeneracy at $\Gamma$, characterized by the $\mathbf{k}$-dependent eigenergies
\begin{equation}
    \varepsilon_n(\mathbf{k})
    =
    \frac{k^2}{2 m^*_n (\hat{\mathbf{k}})}
\end{equation}
and normalized direction-dependent eigenstates represented as three-component vectors
\begin{equation}\label{eq:lk-wf}
\vec{\psi}_{n}(\hat{\mathbf{k}})
=
\begin{pmatrix}
    \psi_x \\
    \psi_y \\
    \psi_z
\end{pmatrix}_n(\hat{\mathbf{k}})
.
\end{equation}
In the cubic generalized Fr\"ohlich model, the former quantities define the electronic term in Eq.~(\ref{gfr:el}). They are modified by the $\sigma$ parameter, which controls the character of band energies (conduction or valence) and the kind of polaron formation (electron or hole).
The latter quantities, in turn, enter the definition of $s$-matrices in Eq.~(\ref{gfr:smatrix}).

Furthermore, the LK Hamiltonian ensures the inherent cubic symmetry of the cubic generalized Fr\"ohlich model.
Specifically, both $\hat{H}^\mathrm{LK} (\mathbf{k})$ and $\hat{H}^\mathrm{gFr}$ commute with all 48 symmetry operations of the full octahedral point group $O_h$, which leave a cube unchanged in three-dimensional space.
However, the presence of band degeneracy may give rise to self-trapped polarons with broken symmetry, a topic that will be explored in greater detail in this work.
Before entering the discussion on the symmetry exhibited by individual polarons, we focus on the general variational framework used to obtain them in the strong coupling regime.

\subsection{Variational Approach} 
\label{sec:varpeq}

The strong coupling limit of the Fr\"ohlich model corresponds to the autolocalization of a charge carrier within the potential well formed by the induced lattice deformation.
In the adiabatic approximation, when lattice fluctuations are neglected, and the charge carrier is assumed to instantly adjust to the induced polarization, the ground-state polaronic solution can be obtained using a variational approach.
This idea, initially employed by Landau and Pekar
\cite{Landau1948}
to investigate electron autolocalization was later utilized by Miyake
\cite{j_miyake_ground_1976, j_miyake_strong-coupling_1975}
to obtain the most accurate asymptotic solution for the standard Fr\"ohlich model.

The same concept also extends to the generalized Fr\"ohlich model, allowing the accurate variational solution to be numerically obtained within the framework of variational polaron equations.
\cite{vasilchenko_variational_2022}
In what follows, we outline the key features of this formalism.
Derived from the methodology of Sio \textit{et al.} for modeling of localized polarons,
\cite{sio_ab_2019, sio_polarons_2019}
the variational polaron equations address the efficient optimization of polaron formation energy in Bloch space.
Similar to the original method from which it is derived, this methodology enables the computation of the polaronic spectrum of a system, i.e. localized polaronic states with their corresponding energies, wavefunctions, and deformation potentials.
However, due to its variational formulation, it facilitates gradient-based optimization techniques, which scale better in contrast to the self-consistent eigenvalue approach of Sio \textit{et al.}\cite{vasilchenko_variational_2022}

Under the adiabatic approximation, the formalism separates the processes of charge localization and lattice deformation, treating them independently.
Charge localization is linked to the electronic part of the polaron wavefunction $\phi$, expressed in a complete basis set of states $\psi_{n\mathbf{k}}$ with energies $\varepsilon_{n\mathbf{k}}$:
\begin{equation}\label{eq:varphi}
    \phi (\mathbf{r}) = \frac{1}{\sqrt{N_p}} \sum_{n\mathbf{k}} A_{n\mathbf{k}} \psi_{n\mathbf{k}} (\mathbf{r}).
\end{equation}
Here, $N_p$ represents the size of the Born-von Karman supercell, defined by the corresponding $\mathbf{k}$-sampling of the BZ. The variational coefficients $A_{n\mathbf{k}}$ adhere to the normalization condition:
\begin{equation}\label{eq:anorm}
     \sum_{n\mathbf{k}} | A_{n\mathbf{k}} |^2 = N_p.
\end{equation}
The treatment of the deformation potential involves the variation of atomic displacements $\Delta \boldsymbol{\tau}$: an atom $\kappa$ with mass $M_\kappa$ in a unit cell $p$ experiences a displacement from its equilibrium position in a direction $\alpha$ by a collective contribution of phonons eigenmodes $e_{\kappa \alpha,\nu} (\mathbf{q})$ with frequencies $\omega_{\mathbf{q}\nu}$:
\begin{equation}\label{deltatau}
    \Delta \tau_{\kappa \alpha p}
    = - \dfrac{2}{N_p} \sum_{\mathbf{q}\nu}
    B^*_{\mathbf{q}\nu} \left(\dfrac{1}{2M_{\kappa}\omega_{\mathbf{q}\nu}}\right)^{1/2}
    e_{\kappa \alpha, \nu}(\mathbf{q}) e^{i\mathbf{q}\cdot\mathbf{R}_p}.
\end{equation}
Here, $B^*_{\mathbf{q}\nu}$ are the variational coefficients accounting for the contribution of each phonon to the displacements.

This results in the self-consistent variational problem within electronic and vibrational subspaces, denoted by $\boldsymbol{A} \equiv \left\{ A_{n\mathbf{k}} \right\}$ and $\boldsymbol{B} \equiv \left\{ B_{\mathbf{q}\nu} \right\}$ space, respectively.
The variational expression for the energy of a polaron is formulated as
\begin{multline}\label{eq:varpeq}
    E_\mathrm{pol} \left( \boldsymbol{A}, \boldsymbol{B}  \right)
    = E_\mathrm{el} \left( \boldsymbol{A} \right)
    + E_\mathrm{ph} \left( \boldsymbol{B}  \right)
    + E_\mathrm{el-ph} \left( \boldsymbol{A}, \boldsymbol{B}  \right)
\end{multline}
where 
\begin{align}
    &
    E_\mathrm{el} \left( \boldsymbol{A} \right)
    = \frac{1}{N_p} \sum_{n\mathbf{k}}
    | A_{n\mathbf{k}} |^2
    \left( \varepsilon_{n\mathbf{k}} -
    \varepsilon \right)
    + \varepsilon, \label{eq:varpeq:el} \\
    &
    E_\mathrm{ph} \left( \boldsymbol{B} \right)
    = \frac{1}{N_p} \sum_{\mathbf{q}\nu}
    | B_{\mathbf{q}\nu} |^2
    \omega_{\mathbf{q}\nu}, \label{eq:varpeq:ph} \\ 
    \begin{split} \label{eq:varpeq:elph}
    & E_\mathrm{el-ph} \left( \boldsymbol{A}, \boldsymbol{B} \right)
    = \\
    & - \frac{1}{N_p^2} \sum_{\substack{mn\nu \\ \mathbf{kq}}}
    A^*_{m\mathbf{\mathbf{k+q}}}
    B^*_{\mathbf{q}\nu} g_{mn\nu}(\mathbf{k, q})
    A_{n\mathbf{\mathbf{k}}} + \mathrm{(c.c)} 
    \end{split}.
\end{align}
Here, $\varepsilon$ is the Lagrange multiplier introduced to take into account the normalization condition of Eq.~(\ref{eq:anorm}), and electronic and vibrational degrees of freedom are connected by the electron-phonon matrix elements $g_{mn\nu}(\mathbf{k, q})$.

The variational problem defined by Eqs.~(\ref{eq:varpeq})-(\ref{eq:varpeq:elph}) can be solved by using any (gradient-based) optimization technique, e.g. preconditioned conjugate-gradient approach.
For a known charge localization $\boldsymbol{A}$ the phonon gradient is always set to zero if the deformation potential $\boldsymbol{B}$ is
\begin{equation}\label{eq:phgrad}
    B_\mathbf{\mathbf{q}\nu} \left( \boldsymbol{A} \right)
    = \frac{1}{N_p}\sum_{mn\mathbf{k}}
    A^*_{m\mathbf{k+q}}
    \frac{g_{mn\nu}(\mathbf{k, q})}{\omega_{\mathbf{q}\nu}}
    A_{n\mathbf{k}}.
\end{equation}
At any system configuration, the Lagrange multiplier is obtained as 
\begin{multline}\label{eq:eps}
    \varepsilon \bigl( \boldsymbol{A}, \boldsymbol{B} \bigr)
    = \frac{1}{N_p}
    \sum_{n\mathbf{k}} | A_{n\mathbf{k}} |^2 \varepsilon_{n\mathbf{k}} \\
    - \frac{1}{N_p^2} \sum_{\substack{mn\nu \\ \mathbf{kq}}}
    \left(
    A^*_{m\mathbf{k+q}}
    B^*_{\mathbf{q}\nu} g_{mn\nu}(\mathbf{k}, \mathbf{q})
    A_{n\mathbf{k}} + (c.c.)
    \right),
\end{multline}
and the gradient with respect to electronic degrees of freedom $A_{n\mathbf{k}} $ reads as
\begin{multline}\label{eq:elgrad}
    D_{n\mathbf{k}} \bigl( \boldsymbol{A}, \boldsymbol{B}, \varepsilon \bigr)
    = \frac{2}{N_p} A_{n\mathbf{k}}
    \left( \varepsilon_{n\mathbf{k}} - \varepsilon \right) \\
    - \frac{2}{N_p^2}
    \sum_{m\nu\mathbf{q}}
    \bigl(
    A_{m\mathbf{k-q}}
    B^*_{\mathbf{q}\nu} g_{nm\nu}(\mathbf{k-q}, \mathbf{q}) \\ 
    +
    A_{m\mathbf{k+q}}
    B_{\mathbf{q}\nu} g^*_{mn\nu}(\mathbf{k}, \mathbf{q})
    \bigr).
\end{multline}

To find a solution, a self-consistent approach is required for these equations. The process begins with an initial set of electronic coefficients $\boldsymbol{A}$, from which the induced deformation field $\boldsymbol{B}$ is determined using Eq.~(\ref{eq:phgrad}).
Subsequently, these values are employed together to calculate the energy of a localized polaronic state $\varepsilon$ based on Eq.~(\ref{eq:eps}).
A step along the steepest-descent direction is then taken using Eq.~(\ref{eq:elgrad}), and this process is iterated until the global minimum of the polaron energy is reached.
This self-consistent procedure reflects the adiabatic character of the problem, where the electronic wavefunction immediately adapts to the deformation, and vice versa, resulting in a self-trapped solution.

The variational equations outlined above are formulated independently of the input parameters, which include band energies  $\varepsilon_{n\mathbf{k}}$, phonon frequencies $\omega_{\mathbf{q}\nu}$ and EPI matrix elements $g_{mn\nu}(\mathbf{k, q})$.
Consequently, any set of parameters can be applied, whether they are fully \textit{ab initio} values obtained from DFT calculations or model input, such as the generalized Fr\"ohlich parametrization.

\subsection{Variational Treatment of the Cubic Generalized Fr\"ohlich Model}
\label{sec:varpeq-cubic}

Variational polaron equations provide a convenient method for investigating the generalized Fr\"ohlich formalism.
In what follows, we introduce the variational framework tailored for the generalized Fr\"ohlich model and cubic systems that exhibit three-fold degeneracy.

As discussed in Section~\ref{sec:lk}, the ground-state configuration of cubic materials with three-fold degeneracy is considered through the LK Hamiltonian.
As the LK eigenstates $\vec{\psi}_{n}(\hat{\mathbf{k}})$ constitute a complete basis, following the variational approach the electronic component of the polaron wavefunction is expressed in terms of them as in Eq.~(\ref{eq:varphi}):
\begin{equation}\label{gfr_cubic:wf}
    \vec{\phi} (\mathbf{r})
    = \frac{1}{\sqrt{N_p}} \sum_{n\mathbf{k}}
    A_{n\mathbf{k}} \vec{\psi}_{n}(\hat{\mathbf{k}}) e^{i\mathbf{k \cdot r}}.
\end{equation}

The $s$-matrices entering the definition of the electron-phonon matrix elements in Eq.~(\ref{gfr:epi}) are derived from Eq.~(\ref{gfr:smatrix}) as a projection of the LK eigenstates on the degenerate states at $\Gamma$:
\begin{equation}\label{eq:lk_smatrix}
    s_{nm}(\hat{\mathbf{k}})
    = 
    \vec{\psi}^*_n (\hat{\mathbf{k}})
    \cdot 
    \vec{\psi}_m (0).
\end{equation}

Furthermore, due to the adiabaticity and the strong coupling regime, the variational treatment allows the coupling to all modes to be captured by the single mode-independent EPI matrix element,\cite{guster_frohlich_2021} and Eq.~(\ref{gfr:epi}) simplifies to
\begin{multline}\label{gfr_cubic:epi*}
     g_{n'n}^\mathrm{gFr}(\mathbf{k, k' - k})
     = \frac{1}{q}
     \left(
      \frac{2\pi\omega^\mathrm{eff}}{N_p \Omega_0}
      \left( {\epsilon^*} \right)^{-1}
      \right)^{1/2}
      \\
      \times \sum_m
      s_{n'm}(\hat{\mathbf{k'}}) s^*_{nm}(\hat{\mathbf{k}}).
\end{multline}
The total contribution from each mode is encompassed by the effective phonon frequency $\omega^\mathrm{eff}$ and effective dielectric constant $\epsilon^*$.
The latter is given as
\begin{equation}\label{eq:eps1}
    \left( {\epsilon^*} \right)^{-1} =
    \sum_\nu
    \left( \epsilon_\nu^* \right)^{-1},
\end{equation}
where
\begin{equation}
    \left( \epsilon_\nu^* \right)^{-1}
    =
    \frac{4\pi}{\Omega_0}
    \left(
    \frac{1}{\varepsilon^\infty}
    \frac{p_{\nu,\mathrm{LO}}}{\omega_{\nu,\mathrm{LO}} } \right)^2,
\end{equation}
and can be obtained from the static and high-frequency dielectric constants \cite{miglio_predominance_2020, Gonze1997}
\begin{equation}\label{eq:epsiloneff}
    \left( {\epsilon^*} \right)^{-1}
    =
    \left( {\epsilon^\infty} \right)^{-1}
    -
    \left( {\epsilon^0} \right)^{-1}.
\end{equation}
The effective phonon frequency, in turn, is deduced from Eq.~(\ref{deltatau}) and reads as
\begin{equation}\label{eq:omegaeff}
    \omega^\mathrm{eff}
    =
    \left(
    \epsilon^*
    \sum_\nu
    \left( \epsilon_\nu^* \right)^{-1}
    \omega_{\nu,\mathrm{LO}}^{-2}
    \right)^{-1/2}.
\end{equation}
The derivation of Eq.~(\ref{eq:omegaeff}), along with an alternative strategy for determining the effective phonon frequency, is provided in Appendix~\ref{app:omegaeff}.

From Eqs.~(\ref{eq:phgrad}) and (\ref{gfr_cubic:epi*}) it can be shown that all the terms in the variational expression given by Eq.~(\ref{eq:varpeq}) are frequency-independent.
Hence, the specific value of effective phonon frequency $\omega^\mathrm{eff}$ does not impact the variational solution, a consequence of the adiabatic nature of the problem.

Combining Eqs.~(\ref{gfr_cubic:wf}) and (\ref{gfr_cubic:epi*}) with the variational formalism outlined in the previous section, the expression for the polaron formation energy in the generalized Fr\"ohlich model with cubic symmetry is given by (we set $\omega^\mathrm{eff} = 1$ for convenience):
\begin{multline}\label{gfr:variational*}
    E^\mathrm{gFr}_\mathrm{pol} \left( \boldsymbol{A}, \boldsymbol{B}  \right)
    =
    \frac{1}{N_p} \sum_{n\mathbf{k}}
    | A_{n\mathbf{k}} |^2
    \left( \varepsilon_{n}(\mathbf{k}) - \varepsilon \right)
    + \frac{1}{N_p} \sum_{\mathbf{q}}
    | B_{\mathbf{q}} |^2
    \\
    - \frac{1}{N_p^2} \sum_{\substack{mn \\ \mathbf{kk'}}}
    A^*_{m\mathbf{k'}}
    B^*_{\mathbf{k' - k}} g^\mathrm{gFr}_{mn}(\mathbf{k, k' - k})
    A_{n\mathbf{\mathbf{k}}} + \textrm{(c.c)}.
\end{multline}
The exact variational solution can be obtained through a gradient-based minimization as sketched in 
Eqs.~(\ref{eq:phgrad})-(\ref{eq:elgrad}).

Optimization of Eq.~(\ref{gfr:variational*}) provides a spectrum of localized solutions, which are the fixed points of the iterative optimization process.
They are characterized by the electronic wavefunction $\boldsymbol{A}$ and deformation field $\boldsymbol{B}$ that gives zero-gradient in Eq.~(\ref{eq:elgrad}) and result in different formation energies $E_\mathrm{pol}$.
Depending on the value of $E_\mathrm{pol}$, one can obtain either ground-state or higher-energy localized polarons.
Additionally, each polaronic solution can be distinguished by its symmetry.
In the subsequent discussion, we analyze the connection between the original cubic symmetry of the model and the potential symmetries exhibited by polaronic solutions.

\subsection{Symmetry of Polarons}
\label{sec:sym}

As mentioned in Section~\ref{sec:lk}, both LK and cubic generalized Fr\"ohlich Hamiltonians commute with symmetry operations of the cubic point group.
Consequently, the associated variational expression shares the same symmetry and retains its structure under any of the 48 transformations belonging to the full octahedral point group $O_h$.
Parameters of Eq.~(\ref{gfr:variational*}) remain invariant under such transformations, implying the possibility of obtaining a variational solution with cubic symmetry.
However, there is a possibility for spontaneous symmetry-breaking. 
Due to the band degeneracies included in the model, the actual ground-state polaron may possess a lower symmetry compared to the initial $O_h$ point group of the Hamiltonian.

To distinguish between the symmetries of different polarons, obtained with the variational formalism, we relate the symmetry of a polaron to the density of its charge localization:
\begin{equation}\label{eq:denstiry}
    \rho (\mathbf{r}) = |\vec{\phi}(\mathbf{r})|^2.
\end{equation}
In essence, a polaron solution belongs to a point group $G$ if, for any symmetry transformation $\hat{S} \in G$, its density remains invariant under the transformation $\mathbf{r} \rightarrow \mathbf{r}' = \hat{S} \mathbf{r}$:
\begin{equation}
    \rho(\mathbf{r}') = \rho (\mathbf{r}).
\end{equation}
In general, an individual polaron solution may belong to one of the 25 subgroups of $O_h$.
Nonetheless, it is possible to limit the possible point groups associated with polarons by analyzing the variational Fr\"ohlich framework used for their calculation.

To analyze the symmetry of generalized Fr\"ohlich polarons obtained with variational formalism in cubic systems, we rely on the symmetry group of deformation potential $\boldsymbol{B}$.
We notice its connection with the charge localization density $\rho$, representative of the actual polaronic symmetry, by rewriting the expression for vibrational coefficients.
Using the Fourier transform of the electronic component of polaron wavefunction in Eq.~(\ref{gfr_cubic:wf}), the generalized Fr\"ohlich EPI in Eq.~(\ref{gfr_cubic:epi*}) and the explicit overlap between LK eigenstates in $s$-matrices in Eq.~(\ref{eq:lk_smatrix}), the vibrational coefficients of Eq.~(\ref{eq:phgrad}) are given by
\begin{equation}\label{eq:bsym}
    B_{\mathbf{q}}
    =
    \frac{g^\mathrm{Fr} (\mathbf{q})}{\omega^\mathrm{eff}}
    \frac{1}{N_p}
    \sum_{\mathbf{k}}
    \vec{\phi}^*(\mathbf{k + q}) \cdot \vec{\phi}(\mathbf{k}).
\end{equation}
Here, $g^\mathrm{Fr} (\mathbf{q})$ corresponds to the overlap-independent part of Eq.~(\ref{gfr_cubic:epi*}), i.e. the EPI of the standard Fr\"ohlich model in the Born and Huang convention
\cite{Born54}:
\begin{equation}\label{gfr:epi*}
     g^\mathrm{Fr}(\mathbf{q})
     = \frac{1}{q}
     \left(
      \frac{2\pi\omega^\mathrm{eff}}{N_p \Omega_0}
      \left( {\epsilon^*} \right)^{-1}
      \right)^{1/2}.
\end{equation}
In the infinite limit of $N_p \to \infty$ the summation on the right-hand side becomes a convolution corresponding to the charge localization density in reciprocal space
\begin{equation}\label{eq:rho-def}
    \rho(\mathbf{q})
    =
    \frac{1}{(2\pi)^{3/2}}
    \int d\mathbf{k}~\vec{\phi}^*(\mathbf{k - q}) \cdot \vec{\phi}(\mathbf{k}).
\end{equation}
Thus, in the infinite limit Eq.~(\ref{eq:bsym}) becomes
\begin{equation}\label{eq:bsym*}
    B_{\mathbf{q}}
    =
    \frac{g^\mathrm{Fr} (\mathbf{q})}{\omega^\mathrm{eff}}
    (2\pi)^{3/2}
    \rho(-\mathbf{q}).
\end{equation}
Since any isometry preserves distances, $g^\mathrm{Fr} (\mathbf{q})$ remains invariant under a transformation $\hat{S}$ such as $\mathbf{q} \rightarrow \mathbf{q}' = \hat{S} \mathbf{q}$.
Therefore, if for any $\hat{S}$ the deformation field remains unchanged $B_\mathbf{q} = B_\mathbf{q'}$, this automatically implies that the same symmetry is applied to the charge localization density $\rho (-\mathbf{q}) = \rho (-\mathbf{q'})$ and vice versa.

Furthermore, the Fr\"ohlich variational formalism implies translation invariance.
From the definition of deformation field coefficients in Eq.~(\ref{eq:phgrad}) and the variational expression in Eq.~(\ref{gfr:variational*}), it can be seen that if a certain charge localization $\boldsymbol{A}$ yields a solution, the same solution is achieved with a phase shift $\left\{ e^{i\mathbf{k}\cdot\mathbf{R}_0}A_{n\mathbf{k}} \right\}$, which corresponds to a $\mathbf{r} \rightarrow \mathbf{r} + \mathbf{R}_0$ translation in real space.

The identical solution for the self-consistent eigenvalue problem, posed by Eqs.~(\ref{eq:phgrad}), (\ref{eq:elgrad}), is also obtained at $\boldsymbol{A}^*$ in Fr\"ohlich case.
This eigenvalue problem corresponds to the steepest-descent optimization in the variational formalism and a linear combination of $\boldsymbol{A}$ and $\boldsymbol{A}^*$ can be taken as a solution, allowing electronic coefficients to be chosen as real-valued.
This results in a real-valued deformation potential $B^*_{\mathbf{q}} = B_{\mathbf{q}}$. Utilizing Eqs.~(\ref{eq:rho-def}) and (\ref{eq:bsym}), this leads to the inversion symmetry invariance for polarons:
\begin{equation}
    \rho(\mathbf{q}) = \rho(-\mathbf{q}).
\end{equation}
Consequently, for degenerate systems, out of the 25 possible subgroups of the $O_h$ point group, only 9 subgroups that exhibit inversion symmetry are allowed: $C_i$, $C_{2h}$, $C_{4h}$, $D_{2h}$, $D_{4h}$, $D_{3d}$, $S_6$, $T_h$, and $O_h$ itself.

However, if a set of model parameters corresponds to a non-degenerate system, the actual symmetry group will be continuous. 
In the case when the generalized Fr\"ohlich model is reduced to the standard Fr\"ohlich model, the resulting polarons will belong to the orthogonal group $O(3)$, representing the full symmetry of a sphere.
If the generalized Fr\"ohlich model describes an anisotropic non-degenerate system, the polarons will possess the $O(2)$$\times$$O(1)$ group of a spheroid.
These scenarios are possible if the coupling between LK bands is removed by setting the $c$ parameter of the LK Hamiltonian to 0.

Lastly, Eq.~(\ref{eq:bsym*}) provides a convenient way to impose a certain symmetry on polaronic solutions within the variational framework.
By attributing a particular symmetry to the deformation field $B_\mathbf{q}$, the same symmetry is automatically applied to the charge localization density of a polaron, $\rho(\mathbf{q})$.
Practically, throughout the minimization process, this is achieved by calculating $B_\mathbf{q}$ explicitly with Eq.~(\ref{eq:phgrad}) only within a symmetry-irreducible wedge of the reciprocal space.
Outside this region, it is reconstructed through the corresponding symmetry operations.
This technique allows us to guide the variational minimization towards a solution possessing one of the 9 possible point groups and explore the polaronic spectrum of a system to indicate ground-state and higher-energy self-trapped polarons.

\section{Computational Details}
\label{sec:setup}

In the current work, the cubic generalized Fr\"ohlich model is solved variationally within the Variational Polaron Equations module implemented in the ABINIT software package.
\cite{gonze_abinitproject_2020, Romero2020}
Self-trapped polaron solutions are obtained using the iterative preconditioned conjugate gradient approach, applied to Eqs.~(\ref{eq:phgrad})-(\ref{eq:elgrad}) to optimize the variational expression in Eq.~(\ref{gfr:variational*}) as detailed in Ref.~\citenum{vasilchenko_variational_2022}.

The input parameters for the model are the LK parameters $a$, $b$, $c$, and effective permittivity $\epsilon^*$.
Since the actual value of the effective phonon frequency entering the variational equations does not affect the results, $\omega^\mathrm{eff}$ is set to 1 during the calculations.

For each set of the parameters, calculations are performed using BvK supercells of increasing size $N_p$.
The size of a BvK cell is determined by the density of the corresponding $\mathbf{k}$-grid, sampling the first BZ.
The polaron formation energy $E_\mathrm{pol}$ in the infinite-size limit $N_p \rightarrow \infty$ is obtained through Makov-Payne extrapolation \cite{Makov1995}
\begin{equation}
    E_\mathrm{pol}(N_p) =
    E_\mathrm{pol}^\infty + \gamma N_p^{-1} + \mathcal{O}(N_p^{-3}).
\end{equation}
For each solution, we also analyze the resulting charge localization $\boldsymbol{A}$, deformation potential $\boldsymbol{B}$, and polaronic density $\rho(\mathbf{r})$.
The density of the $\mathbf{k}$-grid defining the corresponding BvK supercell may vary based on the degree of anisotropy in a system, captured by the ratio between $a$, $b$, and $c$ parameters of the LK Hamiltonian. 
For instance, a $20\times20\times20$ $\mathbf{k}$-grid is sufficient for moderately anisotropic systems to obtain reasonable solutions, but highly anisotropic systems may require denser $\mathbf{k}$-grids, such as $40\times40\times40$.

To facilitate the convergence of the polaronic wavefunction with respect to $N_p$, the divergent electron-phonon matrix elements defined in Eq.~(\ref{gfr_cubic:epi*}) are corrected as
\begin{equation}\label{eq:epi-corrected}
    {g}_{n'n}^\mathrm{gFr}(\mathbf{k}, 0)
    =
    \dfrac{\sqrt{3}}{2\pi}
    \left(
    \dfrac{4\pi N_p \Omega_0}{3}
    \right)^{1/3}
    \left(
    \dfrac{2\pi \omega^\mathrm{eff}}{N_p \Omega_0}
    {(\epsilon^{*})}^{-1}
    \right)^{1/2}
\end{equation}
in the infrared limit.
Eq.~(\ref{eq:epi-corrected}) has the same form and derivation as in the non-degenerate band case, \cite{vasilchenko_variational_2022}
since at $\Gamma$-point the band mixing part matrix elements in Eq.~(\ref{gfr_cubic:epi*}) become diagonal with band index:
\begin{equation}
    \lim_{\mathbf{q} \rightarrow 0}
    \sum_m
    s_{n'm}(\hat{\mathbf{k'}})
    s^*_{nm}(\hat{\mathbf{k}})
    =
    \delta_{n'n}.
\end{equation}

Additionally, since the LK Hamiltonian is non-periodic, the full electronic $\mathbf{k}$-space in variational equations may span beyond the first BZ.
Hence, for a BvK supercell represented by a BZ with sampling $\mathbf{k}_\mathrm{BZ}$, the total $\mathbf{k}$-space is expanded as
\begin{equation}
    \mathbf{k} = \mathbf{k}_\mathrm{BZ} + \mathbf{G},   
\end{equation}
where $\mathbf{G}$ are all the reciprocal space vector shifts.
To ensure its finiteness, the $\mathbf{k}$-space is bounded by the plane wave energy cutoff $\varepsilon_\mathrm{cut}$ such as $\varepsilon_n(\mathbf{k}) \leq \varepsilon_\mathrm{cut}$.
On the other hand, phonon vectors $\mathbf{q}$ = $\mathbf{k'} - \mathbf{k}$ are allowed to extend beyond the cutoff sphere.
During the optimization, the plane wave energy cutoff $\varepsilon_\mathrm{cut}$ is chosen individually for each parameter set, ensuring that the $\mathbf{k}$-space is large enough to accommodate the polaronic wavefunction $\vec{\phi}(\mathbf{k})$.
Convergence with respect to $\varepsilon_\mathrm{cut}$ is straightforward and can be achieved at relatively small $\mathbf{k}$-grids.

In order to compare our results with the already published Gaussian trial wavefunction solution, we take this data from Ref.~\citenum{guster_frohlich_2021} for 19 cubic compounds.
These parameters were obtained within the ABINIT package using GGA-PBE functional \cite{Perdew1996} with the corresponding norm-conserving pseudopotentials from the PseudoDojo project.
\cite{vanSetten2018}
The specific values are listed in Table~S1 of Supplementary Information. \cite{SI}

Except for the aforementioned benchmark case of 19 systems, we do not limit ourselves to a particular set of parameters.
We investigate polaronic solutions across the entire range of $a$, $b$, and $c$ input parameters, encompassing scenarios that yield qualitatively different polaronic solutions of distinct symmetry groups.
To differentiate between possible point groups of polarons, the variational process is directed towards a solution with certain symmetry using Eq.~(\ref{eq:bsym}).
By enforcing a particular symmetry on the deformation potential $\boldsymbol{B}$ during the minimization, we automatically obtain a polaron of respecting the symmetries of the corresponding point group.

\section{Results and Discussion}
\label{sec:results}

The ground-state electronic configuration of a triply degenerate cubic system is entirely determined by the LK Hamiltonian parameters $a$, $b$, and $c$.
In this context, three distinct scenarios can be outlined, namely two simple special cases, and the general case:
(i) $a = b$, $c \equiv 0$ -- three fully identical isotropic bands, which corresponds to the isotropic non-degenerate case (standard Fr\"ohlich model for each of the bands);
(ii) $a \neq b$, $c \equiv 0$ -- three uniaxial bands, with the ground-state corresponding to a non-degenerate case (anisotropic Fr\"ohlich model for each of the bands);
(iii) $c \neq 0$ -- anisotropic triply-degenerate case.
In this section, we apply the variational treatment of the generalized Fr\"ohlich model to these scenarios and analyze the polaronic solutions.

\subsection{Standard Fr\"ohlich Model}

When $a = b$ and $c \equiv 0$, the LK Hamiltonian is fully isotropic:
\begin{equation}
    \hat{H}_{ij}^\mathrm{LK} (\mathbf{k})
    =
    a \left( k_x^2 + k_y^2 + k_z^2 \right) \delta_{ij},
\end{equation}
yielding three spherical bands that are degenerate across the entire $\mathbf{k}$-space.
In the effective mass representation, these bands are expressed as:
\begin{equation}
    \varepsilon_i (\mathbf{k})
    =
    \frac{1}{2m^*}
    \left(  k_x^2 + k_y^2 + k_z^2 \right),
\end{equation}
where the isotropic effective mass is $m^* = 0.5 a^{-1}$.
During polaron formation, each of the three bands can contribute to charge localization, and this triplet can be considered as a single isotropic band in this context.
This aligns precisely with the standard Fr\"ohlich model, for which an asymptotic solution is known.
\cite{j_miyake_strong-coupling_1975, j_miyake_ground_1976}
In the adiabatic strong-coupling regime, this solution is given by
\begin{equation}\label{eq:fr-sol}
    E_\mathrm{pol} = -0.1085 \alpha^2 \omega^\mathrm{eff}, 
\end{equation}
where $\alpha$ is a dimensionless coupling constant
\begin{equation}
    \alpha
    =
    \left(
    \frac{m^*}{2\omega^\mathrm{eff}}
    \right)^{1/2}
    (\epsilon^*)^{-1}.
\end{equation}

By setting $a = b = 0.25$, $c = 0$ we initialize a system with three identical spherical bands, corresponding to the standard Fr\"ohlich model with $m^*$~=~2.
Choosing the effective permittivity $\epsilon^* = 1$ and solving Eq.~(\ref{gfr:variational*}) variationally we obtain the value of polaron formation energy $E^\mathrm{gFr}_\mathrm{pol}$~=~$-0.1084$~Ha.
This is only 0.01~\% higher than the best asymptotic solution given by Eq~(\ref{eq:fr-sol}), which yields $E_\mathrm{pol}$~=~$-0.1085$~Ha for the considered values of effective mass and permittivity.
The resulting polaronic density $\rho(\mathbf{r})$ is spherical and possesses the orthogonal group $O(3)$.

\subsection{Anisotropic Fr\"ohlich  Model}

Having validated the model in the isotropic scenario corresponding to the standard Fr\"ohlich model, we now consider an anisotropic and non-degenerate case.
When $a \neq b$ and $c \equiv 0$, the LK Hamiltonian takes form:
\begin{multline}
    \hat{H}_{ij}^\mathrm{LK} (\mathbf{k})
    =
    \\
    \begin{bsmallmatrix}
        a k_x^2 + b (k_y^2 + k_z^2) & 0 & 0 \\
        0 & a k_y^2 + b (k_x^2 + k_z^2) & 0 \\
        0 & 0 & a k_z^2 + b(k_x^2 + k_y^2) 
    \end{bsmallmatrix}
\end{multline}
and corresponds to three uniaxial bands with degeneracy at $\Gamma$:
\begin{equation}
    \varepsilon_u (\mathbf{k})
    =
    \frac{1}{2m^*} k_u^2
    + 
    \frac{1}{2m^*_\perp}\left( k_v^2 + k_w^2 \right),
\end{equation}
where $u$, $v$, and $w$ indices are used to denote distinct Cartesian coordinates.
Each elliptic band possesses an out-of-plane effective mass $m^* = 0.5 a^{-1}$ along its dedicated Cartesian direction and in-plane effective masses $m^*_\perp = 0.5 b^{-1}$ in the two other directions.

Variational minimization results in a ground-state polaron, with charge localization determined solely by one of the bands, while the other two do not contribute to the process.
This corresponds to the anisotropic Fr\"ohlich model, discussed in Refs.~\citenum{vasilchenko_variational_2022}~and~\citenum{guster_frohlich_2021}.

We recover previously obtained variational solutions from Ref.~\citenum{vasilchenko_variational_2022} but extend them for higher degrees of anisotropy, controlled by the anisotropy parameter $\mu = b/a$ representing the ratio between in-plane and out-of-plane effective masses $m^*/m^*_\perp$.
To benchmark the obtained results, we compare them with the Gaussian trial wavefunction solution from Ref.~\citenum{guster_frohlich_2021} with $a$ fixed to 0.25 (corresponding to $m^* = 2$) and varying the $b$ parameter to control the degree of anisotropy $\mu$.
The effective permittivity $\varepsilon^*$ is set to 1.

\begin{figure}[t]
    \centering
    \includegraphics[scale=0.315]{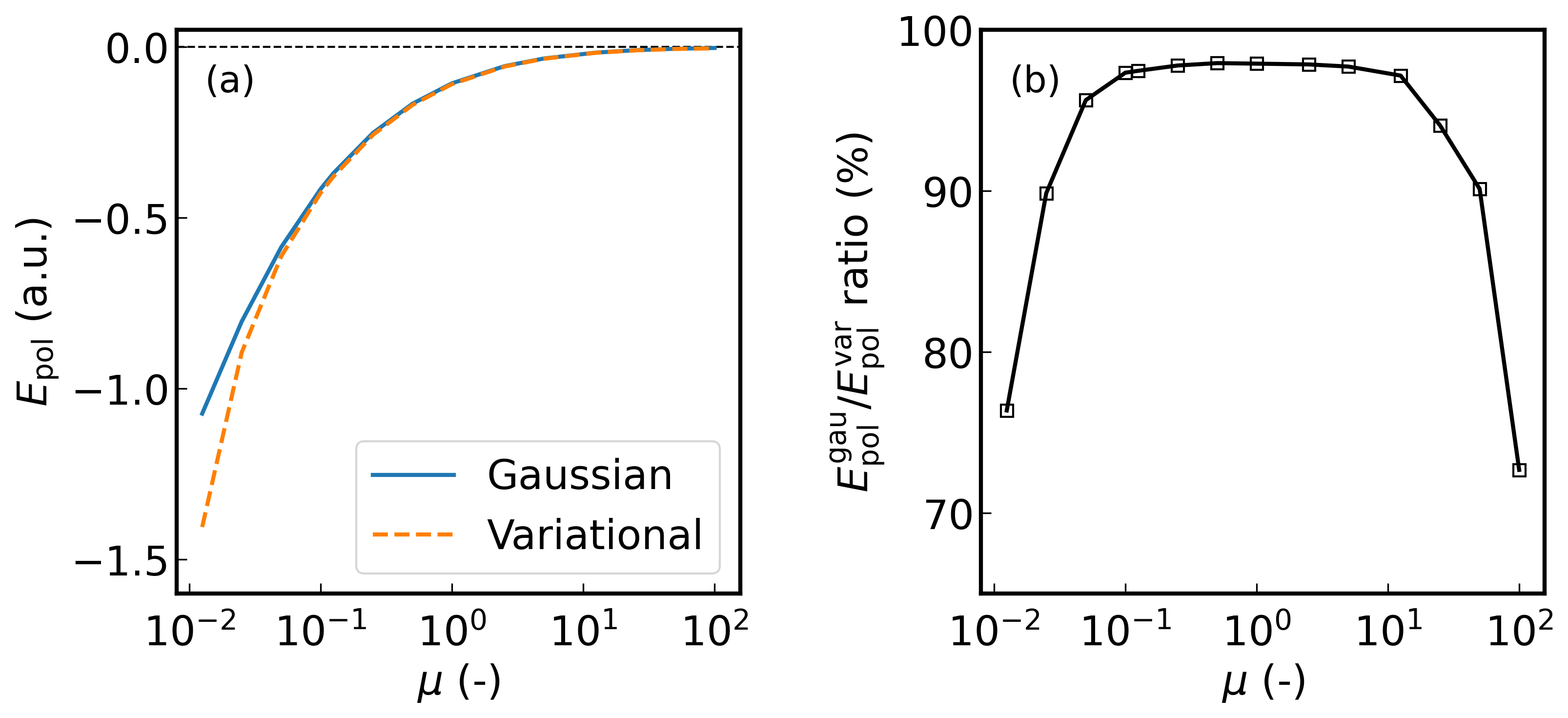}
    \caption{Polaron formation energy $E_\mathrm{pol}$ in anisotropic non-degenerate Fr\"ohlich model, uniaxial case.
    (a) Comparison between $E_\mathrm{pol}$ obtained with Gaussian (blue) and fully variational (orange) approaches.
    (b) Ratio between these energies with respect to the anisotropy parameter $\mu$.
    }
    \label{fig:uniaxial}
\end{figure}

Fig.~\ref{fig:uniaxial} presents the comparison between the Gaussian and fully variational solutions for various degrees of anisotropy $\mu$.
In Ref.~\citenum{vasilchenko_variational_2022}, only the moderate region $10^{-1} < \mu < 10$ was explored, with a relative error of the Gaussian solution not exceeding 3~\%.
However, Fig.~\ref{fig:uniaxial} shows that for higher degrees of anisotropy, the divergence between the Gaussian and the fully variational solution becomes more pronounced, with the variational one yielding up to 30~\% more accurate results.
Hence, Gaussian trial wavefunction becomes inadequate to address polaron formation at high anisotropies and the fully variational treatment is preferred.

\begin{figure}[t]
    \centering
    \includegraphics[scale=0.315]{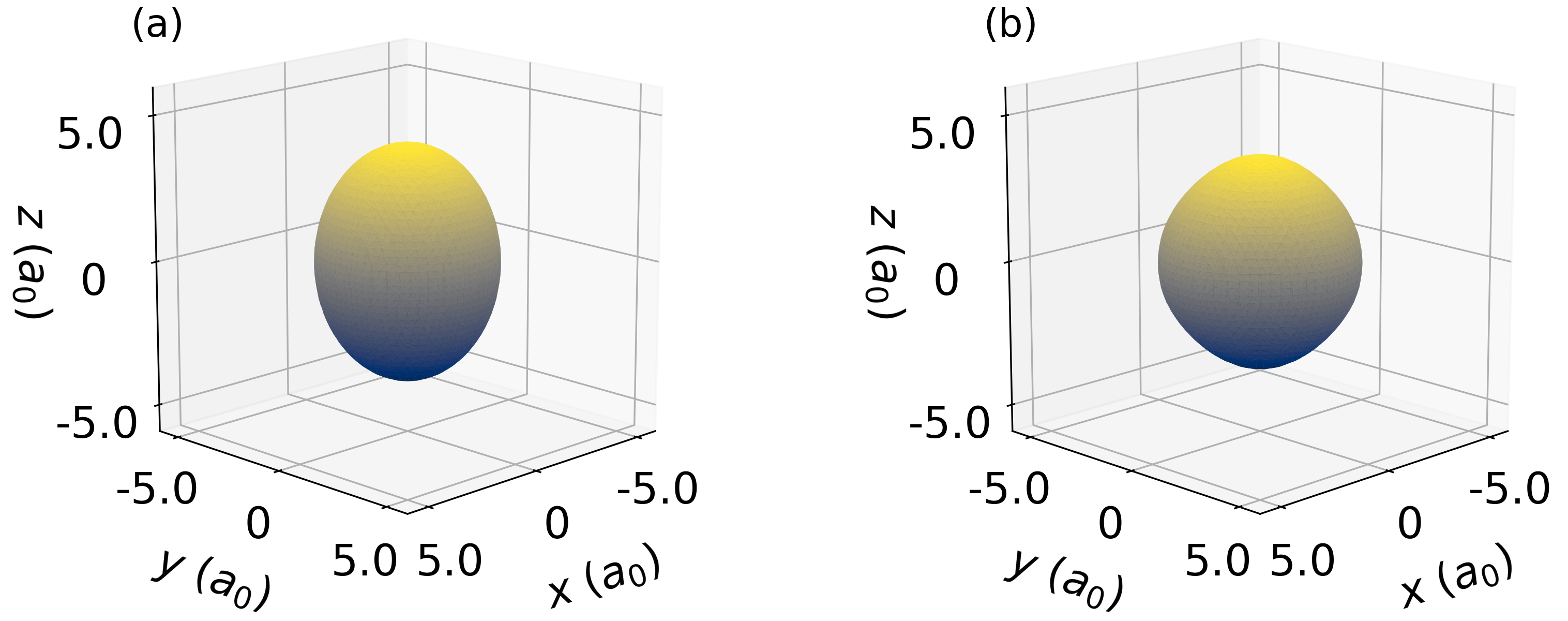}
    \caption{Polaron density surface $\rho(\mathbf{r})$~=~$10^{-4}$ for $a$~=~0.25, $b$~=~0.125, $c$~=~0.
    (a) ground-state, single $\varepsilon_z (\mathbf{k})$ band contribues to the charge localization.
    (b) Higher-energy localized state, all three bands contribute to the charge localization.
    $a_0$ denotes the Bohr radius.}
    \label{fig:uniaxial-den}
\end{figure}

The resulting ground-state polaronic density $\rho(\mathbf{r})$ forms a spheroid,  belonging to the symmetry group $O(2)$$\times$$O(1)$.
This state is triply degenerate since there are three options for choosing a single band $\varepsilon_u (\mathbf{k})$ contributing to the charge localization.
However, during the variational process, it is possible to obtain a higher-energy state, where the three bands equally contribute to the polaron formation.
For instance, with parameters $a$~=~0.25, $b$~=~0.125, $c$~=~0 the polaron formation energies of the ground-state and this solution are  $E_\mathrm{pol}$~=~$-0.1694$~Ha and $E_\mathrm{pol}$~=~$-0.1683$~Ha, respectively.
The comparison between the ground-state spheroid and the higher-energy state density is shown in Fig.~\ref{fig:uniaxial-den}.
To filter out the higher-energy solutions during the optimization process, we encapsulate contribution from a single band $\varepsilon_u (\mathbf{k})$ by choosing an initial charge localization $\boldsymbol{A}$, such as $A_{n\mathbf{k}} \equiv 0$ for $n \neq u$.

\subsection{Cubic Generalized Fr\"ohlich Model}

Setting $c \neq 0$ results in the general form of the LK Hamiltonian given by Eq.~(\ref{eq:h-lk}), which corresponds to the band-degenerate case in the cubic generalized Fr\"ohlich model.
We explore the effects of the band degeneracy on polaronic solutions and, as previously, utilize Gaussian trial wavefunction results of Ref.~\onlinecite{guster_frohlich_2021} for benchmarking.

\begin{figure}[t]
    \centering
    \includegraphics[scale=0.305]{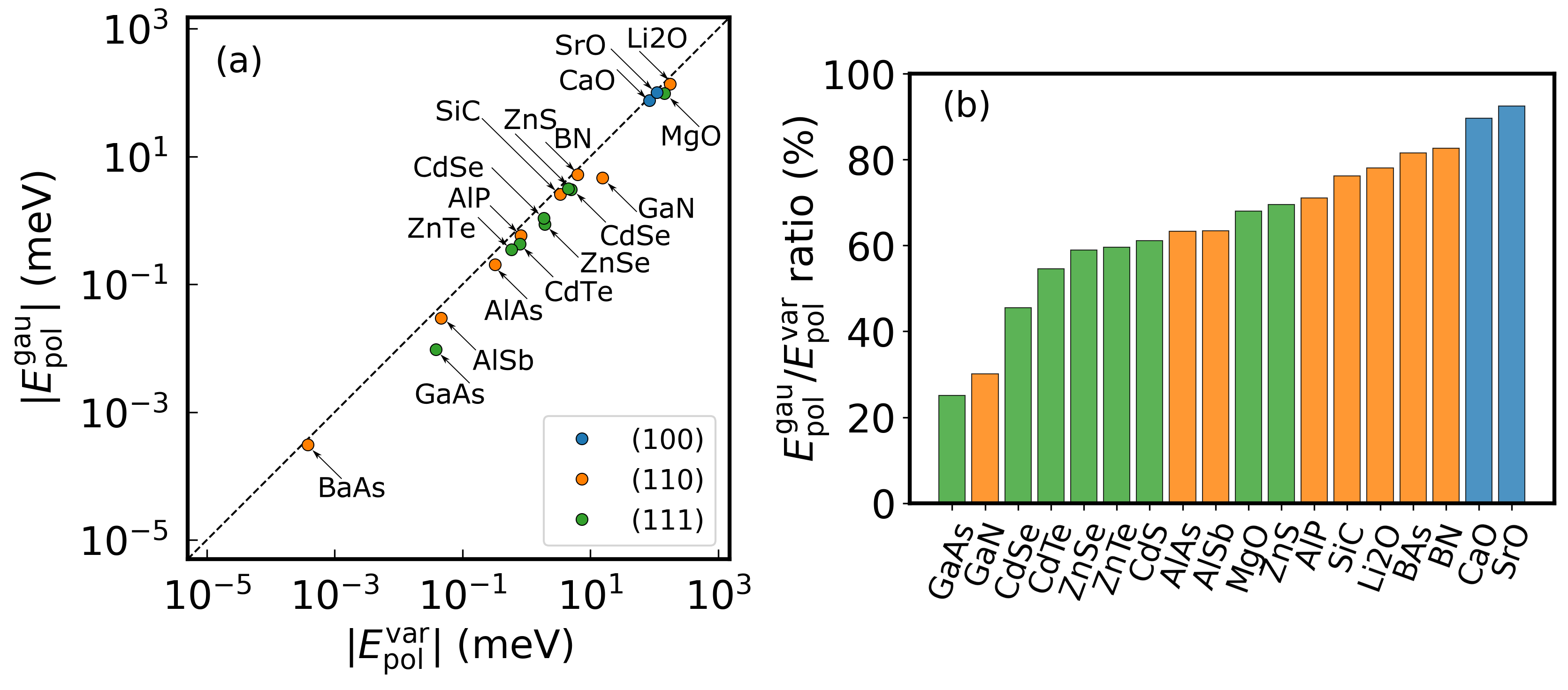}
    \caption{Polaron formation energy $E_\mathrm{pol}$ in the generalized Fr\"ohlich model using LK parameters corresponding to 19 cubic materials, treated without spin-orbit coupling
    (a) Comparison between $E_\mathrm{pol}$ obtained with Gaussian Ansatz and fully variational approaches.
    (b) Ratio between these energies.
    Colors represent the axis of polaron alignment in the Gaussian method.}
    \label{fig:cubic}
\end{figure}

As shown in Fig.~\ref{fig:cubic}, for all of the considered real cubic materials, the fully variational approach outperforms the Gaussian Ansatz, resulting in up to 75~\% more accurate values of the polaron formation energy $E_\mathrm{pol}$ (see Table~S2 of Supplementary Information).\cite{SI}
The reason for that is the unconstrained nature of the variational formalism, with the only requirement being the normalization of the polaronic wavefunction.
On the other hand, in the Gaussian method, one relies on the representation of the LK Hamiltonian as a quadratic form along one of the three symmetry-inequivalent cubic directions in reciprocal space: $[100]$, $[110]$, and $[111]$.
For each direction, the problem is reduced to the non-degenerate case of the anisotropic Fr\"ohlich model,
and the polaron is considered to be aligned with the direction yielding the lowest polaron formation energy $E_\mathrm{pol}$ (represented by the color code in Fig.~\ref{fig:cubic}).

\begin{figure}[t]
    \centering
    \includegraphics[scale=0.315]{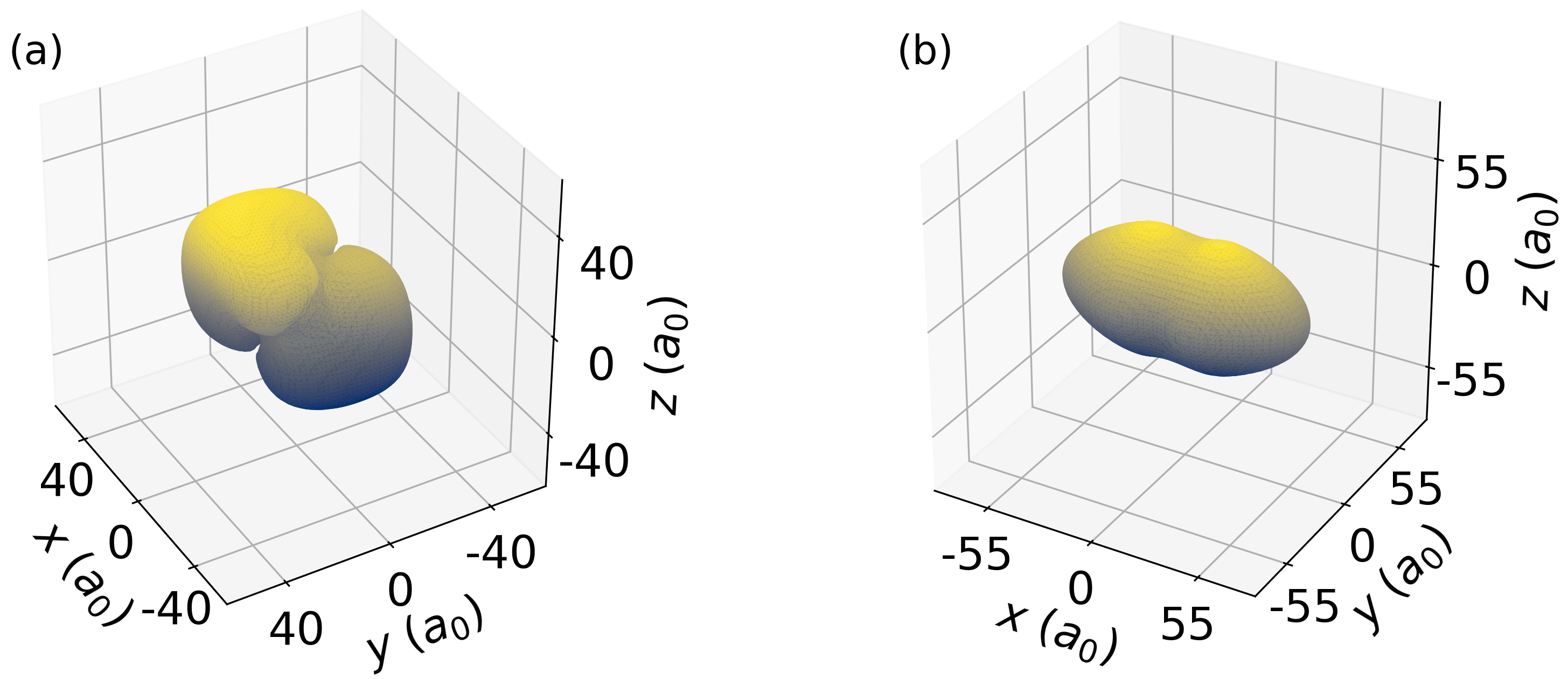}
    \caption{Polaron density surface $\rho(\mathbf{r})$~=~$10^{-8}$ for (a) MgO and (b) CaO treated using the LK Hamiltonian, without spin-orbit coupling.
    The surface value $10^{-8}$ is chosen to display the fine density features.
    $a_0$ denotes the Bohr radius.}
    \label{fig:cubic-den}
\end{figure}

In contrast, the fully variational approach does not have these limitations on the shape and alignment of the polaronic wavefunction, thus giving much more accurate results.
For instance, Fig.~\ref{fig:cubic-den} displays the shapes of the charge localization density $\rho(\boldsymbol{r})$ for polarons
in which the LK Hamiltonian parameters, neglecting SOC, represent those of MgO and CaO obtained with the variational formalism.
For MgO, it resembles a trigonal antiprism with $D_{3d}$ symmetry, while for CaO it is dumbbell-shaped and possesses $D_{4h}$ point group.
While in CaO most of the density can be captured by the Gaussian-shaped trial wavefunction, yielding a 10~\% error, in MgO the fine features of its density result in 32~\% more accurate variational solution.
In both cases, the density is aligned in agreement with the Gaussian Ansatz solution: $[111]$ and $[100]$ for MgO and CaO, respectively.

For other considered materials, in contrast to the Gaussian Ansatz, none of the polarons obtained with the variational formalism aligns with the $[110]$ direction.
Ground-state solutions possess either $D_{3d}$ or $D_{4h}$ point groups with $[111]$ and $[100]$ alignment, respectively.
It is possible to obtain the $[110]$ alignment by enforcing $D_{2h}$ symmetry during the optimization process, which results in higher-energy self-trapped solutions.
In all of the materials, a spontaneous symmetry breaking is observed: starting from the generalized Fr\"ohlich Hamiltonian with $O_h$ cubic symmetry, ground-state polarons of lower symmetries are obtained.

\subsection{Spontaneous Symmetry Breaking of Polarons}

The spontaneous symmetry breaking and the occurrence of a preferred axis for the polaron alignment can be attributed to the band degeneracy present in the model.
To investigate these effects, we systematically analyze symmetry groups of polaronic solutions.
In this case, we do not limit the parameters of the LK Hamiltonian to a particular set of materials and explore them in the wider range of values, limited only by Eqs.~(\ref{eq:lk-limits}), so the Hamiltonian is bounded.

Using the connection between the deformation potential $\boldsymbol{B}$ and charge localization density $\rho({\mathbf{q}})$, discussed in Section~\ref{sec:sym}, we impose certain symmetry groups on polaronic solutions throughout the variational process.
This method allows one, for a particular set of $a$, $b$, and $c$ parameters, to find the ground-state polarons as well as higher-energy localized solutions.
While a system can relax to a ground-state solution without such guidance, this becomes particularly important when the model parameters define a system with two polaronic solutions that are close in energy.
In these calculations, the value of effective permittivity $\epsilon^*$ does not affect the symmetry of solutions and is uniformly set to 1.

To explore the symmetry of ground-state polarons in the wide range of the LK parameters, we first fix the $a$ parameter, and $b$ and $c$ are expressed in units of $a$.
Then, for each value of parameter $b$, $c$ becomes a variable, and the polaron formation energy $E_\mathrm{pol}$ dependence on $c$ is obtained individually for all of the 9 available point groups discussed in Section~{\ref{sec:sym}}.
This procedure allows the exploration of the lowest-energy polarons with the associated point groups and helps to determine the symmetry transition boundaries, i.e. the values of $b$ and $c$ parameters at which the ground-state changes its symmetry.

\begin{figure}[t]
    \centering
    \includegraphics[scale=0.315]{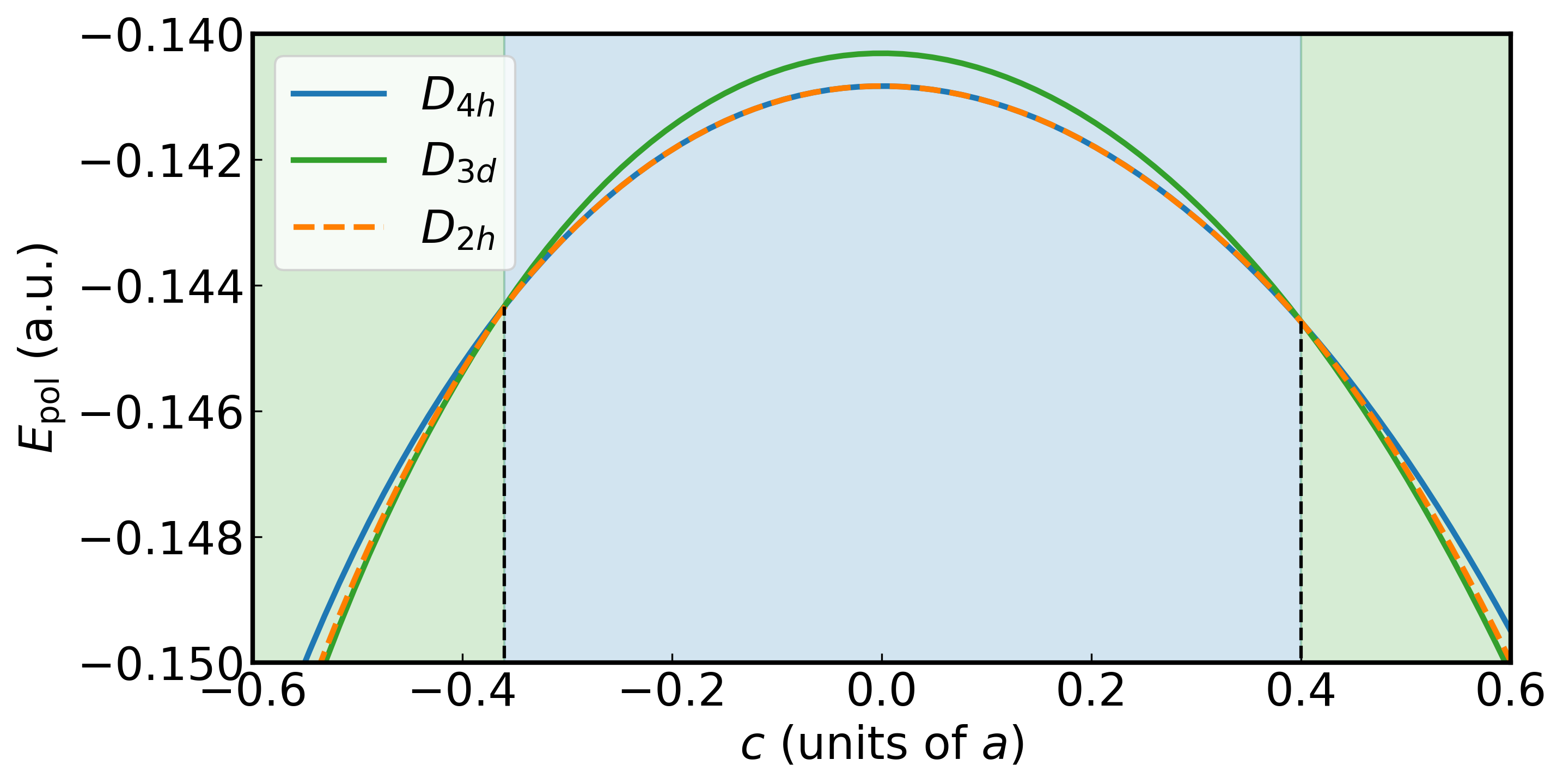}
    \caption{Polaron formation energy $E_\mathrm{pol}$ in the cubic generalized Fr\"ohlich model with respect to $c$ parameter, $a$~=~0.25, $b$~=~0.6$a$.
    Solid blue and green curves, and dashed orange curve represent solutions with $D_{4h}$, $D_{3d}$, and $D_{2h}$ symmetry, respectively.
    Light blue and green shaded regions show the regions of the $D_{4h}$ and $D_{3d}$  
    ground-state solutions, respectively.
    }
    \label{fig:states}
\end{figure}

For instance, for $a$~=~0.25 and $b$~=~0.6$a$, Fig.~\ref{fig:states} presents the dependence of $E_\mathrm{pol}$ on $c$ for solutions with $D_{4h}$, $D_{3d}$ and $D_{2h}$ symmetries.
These three point groups, among the 9 possible ones, lead to self-trapped polarons with the lowest energies across the $c$ range.
Where $D_{4h}$ and $D_{2h}$ result in the same formation energy, the larger $D_{4h}$ group is chosen to represent the ground-state.
Thus, for this choice of parameters, the 
ground-state polaron solution possesses either $D_{4h}$ or $D_{3d}$ symmetry, which is depicted by shaded regions in Fig.~\ref{fig:states}.
The symmetry transition boundaries between these solutions can be easily found at the intersection points of the corresponding $E_\mathrm{pol}(c)$ curves.

\begin{figure}[t]
    \centering
    \includegraphics[scale=0.5]{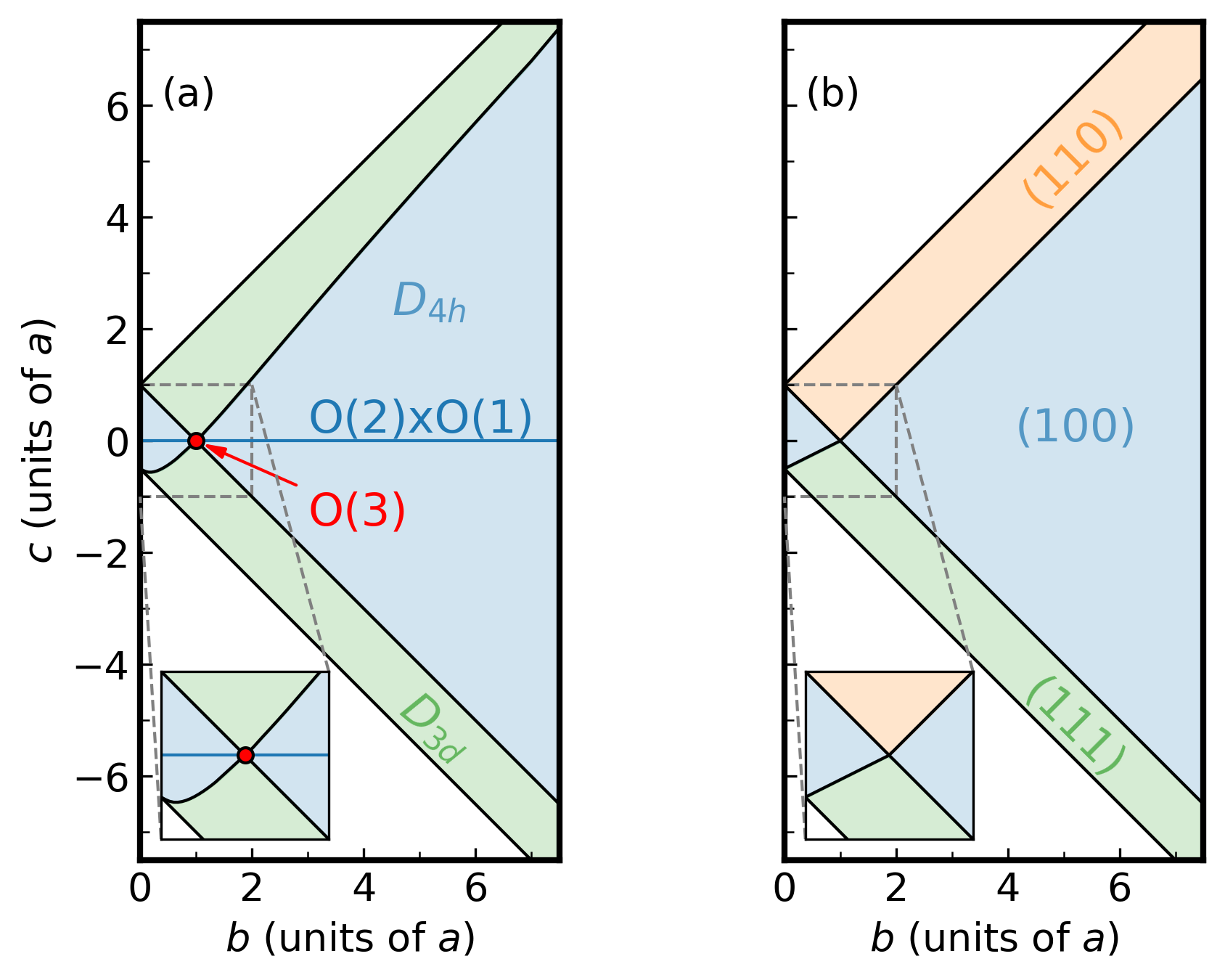}
    \caption{Symmetry phase diagram for polarons in cubic generalized Fr\"ohlich model obtained with (a) fully variational optimization and (b) analysis of effective masses from the LK Hamiltonian.
    Shaded blue, green and orange regions correspond to $D_{4h}$, $D_{3d}$ and $D_{2h}$ point groups, respectively.
    The dark blue line and red circle in panel (a) denote $O(2)$$\times$$O(1)$ and $O(3)$ point groups, respectively, and at these points the problem is reduced to a non-degenerate one.
    Zoomed regions in both panels are used to highlight dissimilarities between the phase diagrams: in panel (a), the lower phase boundary between the $D_{4h}$ and $D_{3d}$ regions is curved, in contrast to panel (b).}
    \label{fig:phase}
\end{figure}

By repeating the aforementioned procedure for a wide range of $b$ parameters, a corresponding symmetry phase diagram is obtained, which is displayed in Fig.~{\ref{fig:phase}}~(a).
It shows that ground-state polarons can exhibit four distinct symmetries depending on the $a$, $b$, and $c$ parameters of the LK Hamiltonian.
While $c \neq 0$ the point group of the lowest-energy solution can be either $D_{4h}$ or $D_{3d}$.
For the trivial case of $c = 0$, as discussed previously, one gets a spheroid with $O(2)$$\times$$O(1)$ symmetry or, if the system is isotropic, a fully spherical polaron of the orthogonal group $O(3)$.
Figures displaying different shapes of charge localization density for each point group in the diagram are provided in the Supplementary Information, Figs. S1--S14.\cite{SI}

In general, systems with higher effective masses are expected to have lower polaron formation energy.
In this sense, polarons are likely to align with the direction of the largest effective mass since this minimizes the kinetic energy contribution in the variational approach.
However, there is a discrepancy between the actual phase diagram and the one that would be obtained from the consideration of effective masses only, without any variational optimization.
In this diagram, shown in Fig.~\ref{fig:phase}~(b), distinct regions correspond to the directions of the largest effective masses, obtained from the LK Hamiltonian by Eqs~(\ref{eq:lk_params*}). They are along the $[100]$, $[110]$ and $[111]$ directions.
In terms of symmetry, these directions would correspond to $D_{4h}$, $D_{2h}$ and $D_{3d}$ point groups.
However, the actual phase diagram shows no sign of the $D_{2h}$ group, and the phase boundaries between the distinct symmetry regions differ slightly. 
Hence a simple analysis of effective masses is not sufficient to deduce the symmetry of a polaron.

However, it is possible to explain the origin of one of the phase boundaries present in both diagrams, from the nature of the LK Hamiltonian: $c = a - b$.
For this purpose, the LK Hamiltonian can be rewritten by splitting it into three parts
\begin{equation}\label{eq:lk_split}
     H^\mathrm{LK} (\mathbf{k})
     =
     k^2 \left(
     b I
     + c |\hat{\mathbf{k}}\rangle \langle \hat{\mathbf{k}} |
     + \left( a - b - c \right) d(\hat{\mathbf{k}})
     \right),
\end{equation}
where $I$ is the identity matrix and
\begin{gather}
    k^2 |\hat{\mathbf{k}}\rangle \langle \hat{\mathbf{k}} |
    =
    \begin{bmatrix}
        k_x^2   & k_x k_y & k_x k_z \\
        k_x k_y & k_y^2   & k_y k_z \\
        k_x k_z & k_y k_z &  k_z^2
    \end{bmatrix},
    \\
    k^2 d(\hat{\mathbf{k}})
    =
    \begin{bmatrix}
        k_x^2 & 0     & 0      \\
        0     & k_y^2 & 0      \\
        0     & 0     &  k_z^2
    \end{bmatrix}.
\end{gather}
In the context of polaron formation, the first two terms in Eq.~(\ref{eq:lk_split}) are axis-independent components, corresponding to the Trebin and R\"ossler model for systems with isotropic degenerate bands.
\cite{Trebin1975}
While these two terms do not yield any preferred orientation for the polaron formation, the third term depends on the choice of axes.
As a result, the change of sign in its prefactor $a - b - c $ results in a change of a polaron orientation and transition between two distinct polaronic symmetries.

Lastly, it is crucial to note how the inclusion of spin-orbit interactions into the model may change the resulting picture of polaron formation.
Specifically, in cubic systems, the interaction arising from spin-orbit coupling alters the degeneracy of electronic states.
This leads to a characteristic 4+2 degeneracy of the valence band maximum, characterized by the degeneracy between two heavy hole and two light hole bands at the $\Gamma$ point, and a downward shift of two split-off bands.

The strength of this shift is quantified by the corresponding split-off energy, $\Delta_\mathrm{SOC}$.
For systems with light atoms, such as oxides, this effect is expected to be weak and can often be neglected in the context of polaron formation.
For example, as shown in Table~S3 of Supplementary Information,\cite{SI} in the set of lighter materials studied, the absolute value of polaron formation energy $E_\mathrm{pol}$ exceeds $\Delta_\mathrm{SOC}$, justifying the approximation.
However, for materials with heavier elements, the change in the character of degeneracy due to the spin-orbit splitting may potentially modify the resulting symmetries and formation energies of ground-state polarons, depending on the value of $\Delta_\mathrm{SOC}$.
In the strong-coupling regime at hand, this effect is uniformly neglected in the literature.
The necessary generalization may be performed by extending the modifications of the weak-coupling formalism of Ref.~\onlinecite{brousseau-couture_effect_2023} into the variational approach.
Nonetheless, the symmetry-breaking effect itself will remain valid irrespective of spin-orbit coupling, as the degeneracy between the heavy and light hole bands persists, which is the driving force for the symmetry-breaking of the model's solutions.

\section{Conclusion}
\label{sec:conclusion}

In the present paper, the generalized Fr\"ohlich model for cubic systems with degenerate bands is investigated in the strong-coupling limit, on the basis of the Luttinger-Kohn three-band electronic Hamiltonian.
Using the variational polaron equations framework and preconditioned conjugate-gradient optimization, \cite{vasilchenko_variational_2022} we obtain fully variational polaronic solutions.
We show that these solutions are more accurate than the previously reported ones, obtained with the Gaussian trial wavefunction approach, \cite{guster_frohlich_2021}, especially in the band-degenerate case.
Moreover, by enforcing certain symmetry on polarons during the variational process, we obtain polaronic spectra of systems: ground-state and higher-energy polarons with their corresponding formation energies $E_\mathrm{pol}$, charge localization $\boldsymbol{A}$, deformation potential $\boldsymbol{B}$ and charge localization density $\rho(\mathbf{r})$.
By analyzing the values of $E_\mathrm{pol}$ for various symmetries and parameters of the model, we obtain the symmetry phase diagram for polarons.
It shows the effect of spontaneous symmetry breaking: starting from the cubic generalized Fr\"ohlich Hamiltonian with inherent full octahedral symmetry $O_h$, ground-state polaronic solutions of lower point groups are obtained when degeneracy is present.
Depending on the $a$, $b$, and $c$ parameters defining the band structure of a cubic system through the LK Hamiltonian, these point groups can be either $D_{4h}$ or $D_{3d}$.
When $c = 0$, the model reduces to the non-degenerate one, and the $O(2)$$\times$$O(1)$ symmetry or the $O(3)$ symmetry are obtained.

As the present work provides a comprehensive analysis of the cubic generalized Fr\"ohlich model in the strong-coupling regime, it may serve as a reference for any forthcoming all-coupling methods treating the same model.
Moreover, it can be extended to explore the point groups of polarons in systems of other symmetries, not only within Fr\"ohlich approximations but also from a fully \text{ab initio} perspective.
Additionally, further generalization of the model through the inclusion of spin-orbit interaction may pave the way for investigating the importance of this effect on polaron formation in cubic systems with heavy elements.

\begin{acknowledgments}

We thank Matteo Giantomassi for help in the ABINIT implementation and Elham Dehghanpisheh for pointing out that the symmetry breaking observed in this study belongs to the class of Jahn-Teller distortions.
V.V. acknowledges funding by the FRS-FNRS Belgium through FRIA.
Computational resources have been provided by the supercomputing facilities of the Universit\'e catholique de Louvain (CISM/UCL), the Consortium des Equipements de Calcul Intensif en F\'ed\'eration Wallonie Bruxelles (CECI) funded by the FRS-FNRS under Grant No. 2.5020.11.

\end{acknowledgments}

\appendix
\renewcommand\thefigure{A\arabic{figure}}

\section{Effective Phonon Frequency} \label{app:omegaeff}

In the variational treatment of the cubic generalized Fröhlich model, we rely on the fact that the overall contribution of individual phonon modes $\omega_{\nu,\mathrm{LO}}$ to the polaron formation can be captured by the coupling to a single effective phonon mode $\omega^\mathrm{eff}$.
In what follows, we provide the necessary explanation for this fact, as well as two distinct ways to define $\omega^\mathrm{eff}$ by mode-averaging.

Firstly, to switch from a multimode basis to an individual mode, one has to define the effective deformation potential as
\begin{equation}
    B^\mathrm{eff}_\mathbf{q}
    = 
    \sum_\nu
    \left( \frac{\omega_{\nu, \mathrm{LO}}}{\epsilon^*_\nu} \right)^{1/2}
    \left( \frac{\epsilon^*}{\omega^\mathrm{eff}} \right)^{1/2}
    B_{\mathbf{q}\nu}.
\end{equation}
With the effective permittivity and electron-phonon matrix elements given by Eqs.~(\ref{gfr_cubic:epi*})-(\ref{eq:epsiloneff}), it is not hard to show that the corresponding vibrational and electron-phonon terms of the variational expression in Eqs.~(\ref{eq:varpeq})-(\ref{eq:varpeq:elph}) become mode-independent:
\begin{align}
    &
    E_\mathrm{ph} \left( \boldsymbol{B} \right)
    = \frac{1}{N_p} \sum_{\mathbf{q}}
    | B^\mathrm{eff}_{\mathbf{q}} |^2
    \omega^\mathrm{eff}, \label{eq:vibeff} \\ 
    \begin{split} \label{eq:elpheff}
    & E_\mathrm{el-ph} \left( \boldsymbol{A}, \boldsymbol{B} \right)
    = \\
    & - \frac{1}{N_p^2} \sum_{\substack{mn \\ \mathbf{kq}}}
    A^*_{m\mathbf{\mathbf{k+q}}}
    {B_\mathbf{q}^\mathrm{eff}}^* g^\mathrm{gFr}_{mn}(\mathbf{k, q})
    A_{n\mathbf{\mathbf{k}}} + \mathrm{(c.c)} 
    \end{split}.
\end{align}
At this stage, $\omega^\mathrm{eff}$ has not yet been defined.
Moreover, its actual value will not change the final result, since there is an invariance in Eqs.~(\ref{eq:vibeff}) and (\ref{eq:elpheff}) under simultaneous rescaling of $\omega^\mathrm{eff}$ and $B^\mathrm{eff}_\mathbf{q}$.
Indeed, with a scaling factor $\gamma$ the transformation
\begin{equation}
\begin{cases}
    \omega^\mathrm{eff} \rightarrow \gamma \omega^\mathrm{eff} \\
    B^\mathrm{eff}_\mathbf{q} \rightarrow \gamma^{-1/2} B^\mathrm{eff}_\mathbf{q}
\end{cases}
\end{equation}
leaves the $E_\mathrm{ph}$ and $E_\mathrm{el-ph}$ invariant, taking into account the rescaling of $g^\mathrm{gFr}_{mn}(\mathbf{k, q})$ in Eq.~(\ref{gfr_cubic:epi*}).
This is a direct consequence of the strong-coupling and adiabatic regime captured by the variational methodology.

Although the specific value of $\omega^\mathrm{eff}$ is irrelevant to the method at hand, it can still be determined starting from two different hypotheses, leading to two different results.

In the first approach, we rely on Eq.~(\ref{deltatau}) for the atomic displacements, demanding that they have to be the same both in the multimode and effective-mode regimes.
This requires us to introduce a normalized effective phonon eigenmode $e^\mathrm{eff}_{\kappa \alpha} (\mathbf{q})$.
With Eqs.~(\ref{gfr_cubic:epi*}) and (\ref{eq:bsym*}) and taking into account that phonons in the generalized Fr\"ohlich model are taken at the zone center, from Eq.~(\ref{deltatau}) after some algebra one obtains
\begin{equation}
    \left(\omega^\mathrm{eff}
    {\epsilon^*}^{1/2} \right)^{-1}
    e^\mathrm{eff}_{\kappa \alpha} (0)
    =
    \sum_\nu
    \left(\omega_\nu
    {\epsilon^*_\nu}^{1/2} \right)^{-1}
    e_{\kappa \alpha, \nu} (0).
\end{equation}
Multiplying the left- and right-hand sides of the equation by their complex conjugate and summing over $\kappa\alpha$ indices, and taking into account the orthonormality of eigenmodes, the expression for the effective phonon frequency is obtained:
\begin{equation}\label{eq:app_omegaeff1}
    \omega^\mathrm{eff}
    =
    \left(
    \epsilon^*
    \sum_\nu
    \left( \epsilon_\nu^* \right)^{-1}
    \omega_{\nu,\mathrm{LO}}^{-2}
    \right)^{-1/2}.
\end{equation}

An alternative strategy would be to go beyond the static regime of the variational formalism.
For this, we follow Ref.~\citenum{de_melo_high-throughput_2023}, which, in turn, relies on the approach of Hellwarth and Biaggio\cite{Hellwarth1999} for the mode averaging.
Introducing the quantity $W_\nu$, which represents the coupling between LO phonon mode $\nu$ and a single electron, the dielectric response of a material is given by
\begin{equation}
    \left( {\epsilon^\infty} \right)^{-1}
    -
    \left( {\epsilon} (\omega) \right)^{-1}
    =
    \sum_\nu
    \frac{W^2_\nu}{\omega^2_{\nu,\mathrm{LO}} - \omega^2}.
\end{equation}
At $\omega$~=~0, taking into account Eqs.~(\ref{eq:eps1})-(\ref{eq:epsiloneff}), after identification of the contribution of each mode, the couplings are expressed as
\begin{equation}\label{eq:W1}
    W^2_\nu = (\epsilon^*_\nu)^{-1} \omega^2_{\nu, \mathrm{LO}}.
\end{equation}
Following Ref.~\citenum{de_melo_high-throughput_2023}, the square of the effective coupling is obtained from the couplings to individual modes as
\begin{equation}
    W^2_\mathrm{eff} = \sum_\nu W^2_\nu
\end{equation}
and the effective phonon frequency is given by
\begin{equation}\label{eq:W3}
    (\omega^\mathrm{eff})^2 = \epsilon^* W^2_\mathrm{eff} .
\end{equation}
Finally, combining Eqs.~(\ref{eq:W1})-(\ref{eq:W3}) one obtains
\begin{equation}\label{eq:app_omegaeff2}
    \omega^\mathrm{eff}
    =
    \left(
    \epsilon^*
    \sum_\nu
    \left( \epsilon_\nu^* \right)^{-1}
    \omega_{\nu,\mathrm{LO}}^{2}
    \right)^{1/2}.
\end{equation}

Note the reciprocity between Eqs.~\ref{eq:app_omegaeff1} and \ref{eq:app_omegaeff2}.
Starting from two different hypotheses, we obtain two different expressions for $\omega_\text{eff}$, which are both valid within the context of the approximations made.
While this distinction may not matter in the purely strong-coupling and adiabatic limit of the problem, the actual choice of $\omega_\text{eff}$ may become important when going beyond this regime.

\newpage
\bibliography{main} 

\end{document}


\title{
Supplementary Information: Polarons in the Cubic Generalized Fr\"ohlich Model: Spontaneous Symmetry Breaking}

\author{Vasilii Vasilchenko}
\affiliation{European Theoretical Spectroscopy Facility, Institute of Condensed Matter and Nanosciences, Universit\'{e} catholique de Louvain, Chemin des \'{e}toiles 8, bte L07.03.01, B-1348 Louvain-la-Neuve, Belgium}
\author{Xavier Gonze}
\affiliation{European Theoretical Spectroscopy Facility, Institute of Condensed Matter and Nanosciences, Universit\'{e} catholique de Louvain, Chemin des \'{e}toiles 8, bte L07.03.01, B-1348 Louvain-la-Neuve, Belgium}

\date{\today}

\maketitle

\onecolumngrid


\section{Supplementary figures}

\begin{figure}[htbp]
    \centering
    \includegraphics[scale=0.5]{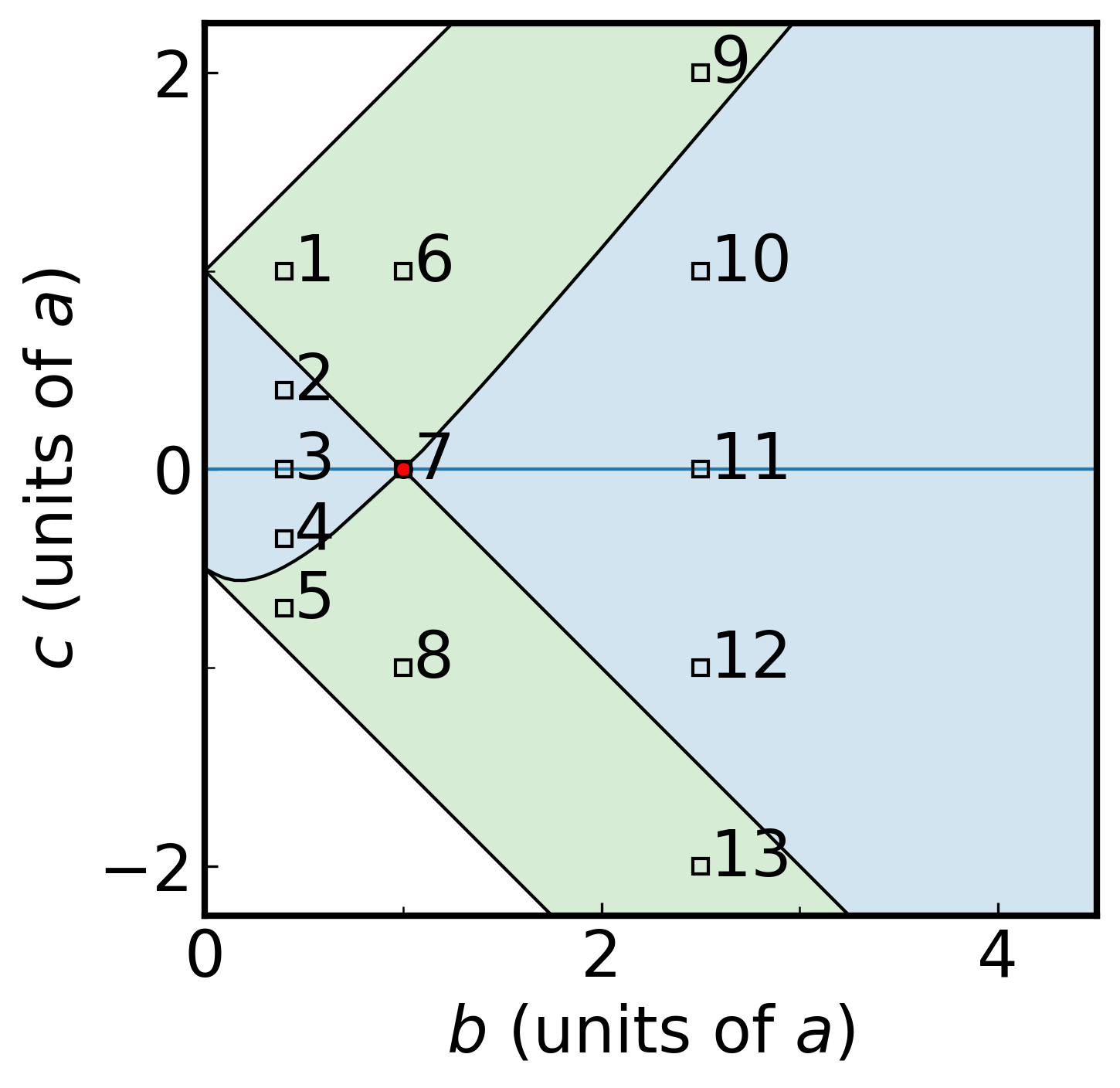}
    \caption{
    Symmetry phase diagram for polarons in cubic generalized Fr\"ohlich model obtained with fully variational optimization.
    Shaded blue and regions correspond to $D_{4h}$ and $D_{3d}$ point groups, respectively.
    The dark blue line and red circle in panel denote $O(2)$$\times$$O(1)$ and $O(3)$ point groups, respectively.
    Numbered markers represent 13 different classes of polarons, shown in Figs.~\ref{fig:den1}--\ref{fig:den13}.
    }
    \label{fig:supp-phases}
\end{figure}

\begin{figure}
    \centering
    \includegraphics[scale=0.5]{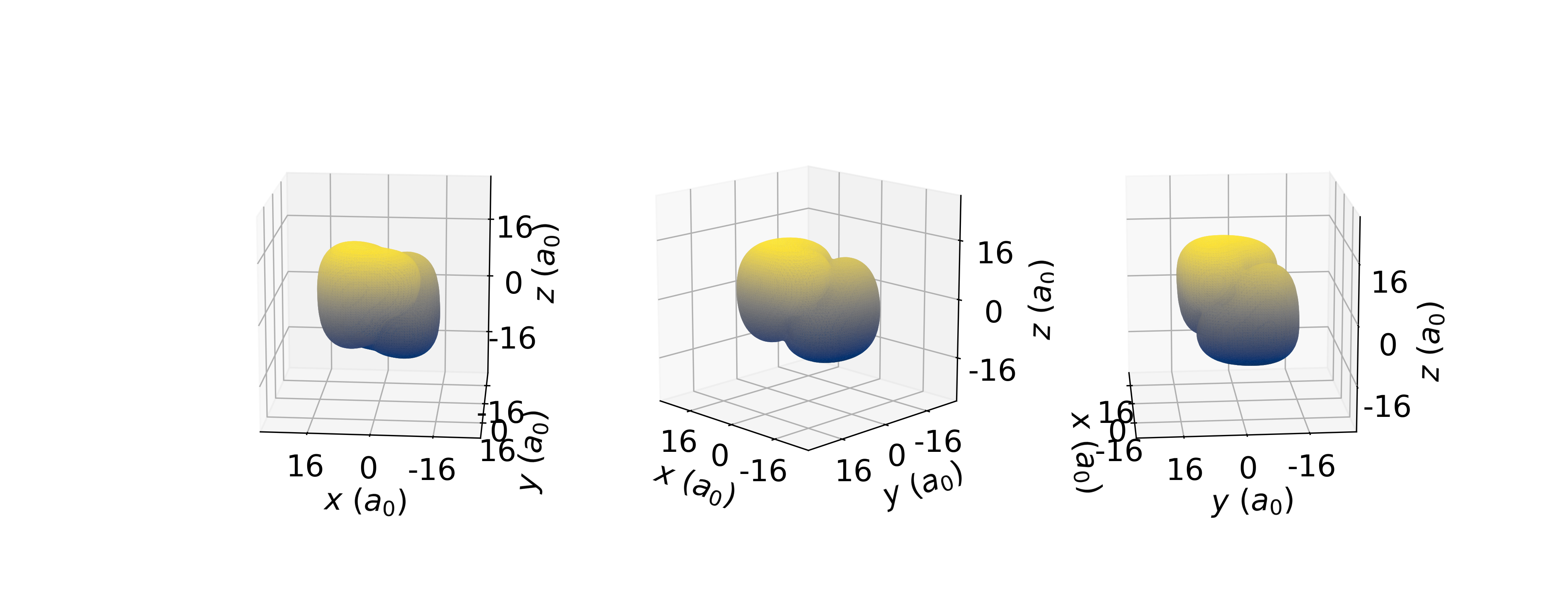}
    \caption{
    1: Polaron density surface $\rho(\mathbf{r}) = 10^{-8}$,
    corresponding to the marker 1 in  Fig.~\ref{fig:supp-phases}.
    $D_{3d}$ point group, $a=1$, $b=0.4$, $c=1.0$}
    \label{fig:den1}
\end{figure}

\begin{figure}
    \centering
    \includegraphics[scale=0.5]{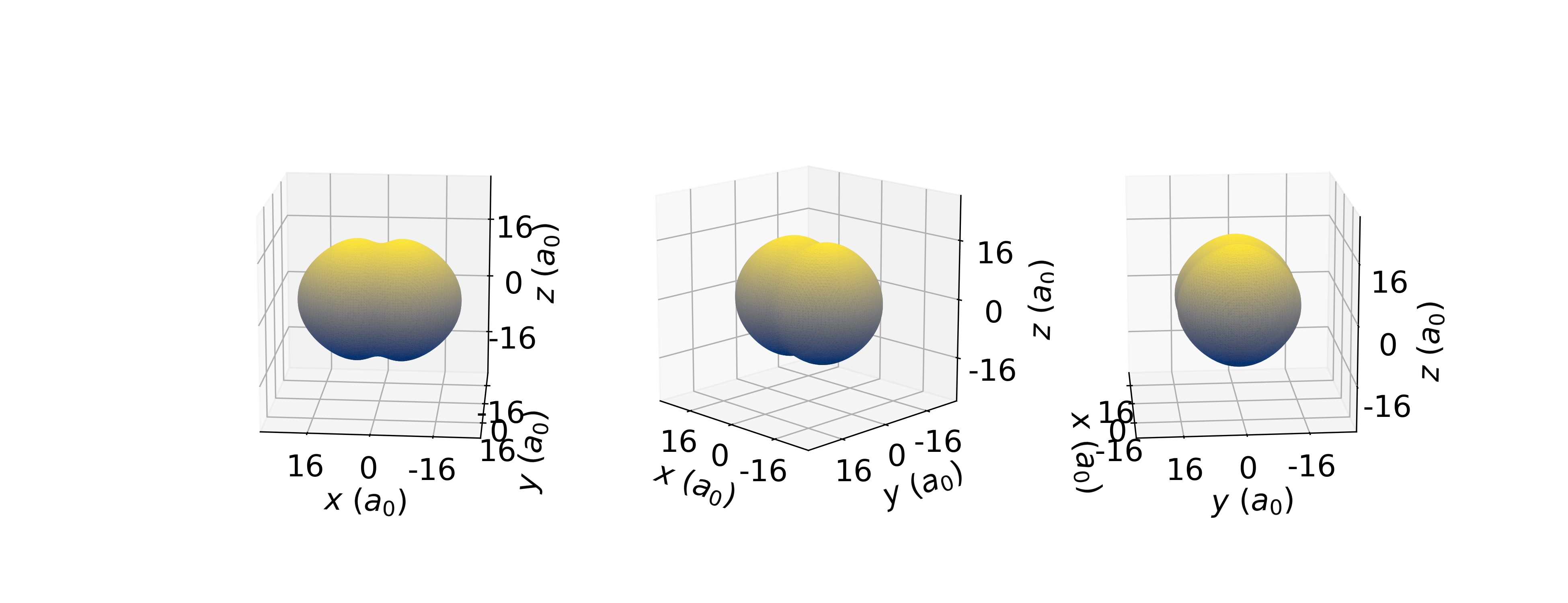}
    \caption{
    2: Polaron density surface $\rho(\mathbf{r}) = 10^{-8}$,
    corresponding to the marker 2 in  Fig.~\ref{fig:supp-phases}.
    $D_{4h}$ point group, $a=1$, $b=0.4$, $c=0.4$}
    \label{fig:den2}
\end{figure}

\begin{figure}
    \centering
    \includegraphics[scale=0.5]{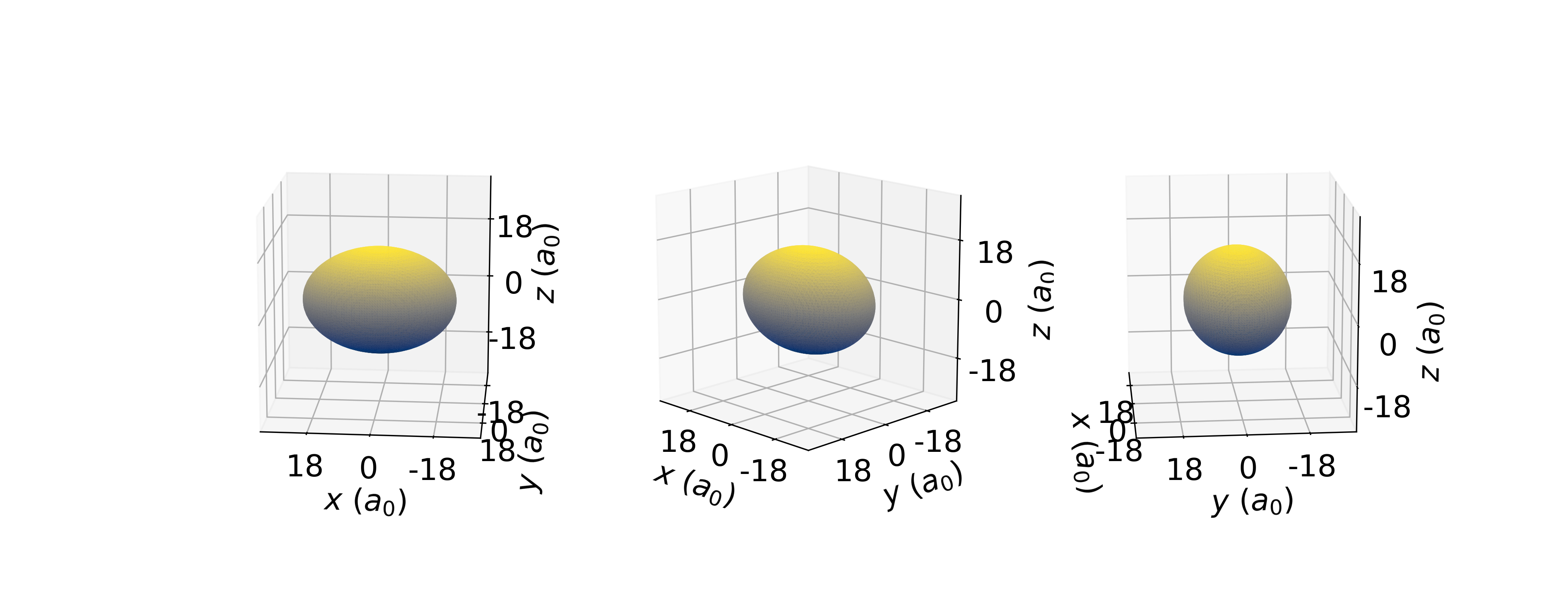}
    \caption{
    3: Polaron density surface $\rho(\mathbf{r}) = 10^{-8}$,
    corresponding to the marker 3 in  Fig.~\ref{fig:supp-phases}.
    $O(2)$$\times$$O(1)$ point group, $a=1$, $b=0.4$, $c=0$}
    \label{fig:den3}
\end{figure}

\begin{figure}
    \centering
    \includegraphics[scale=0.5]{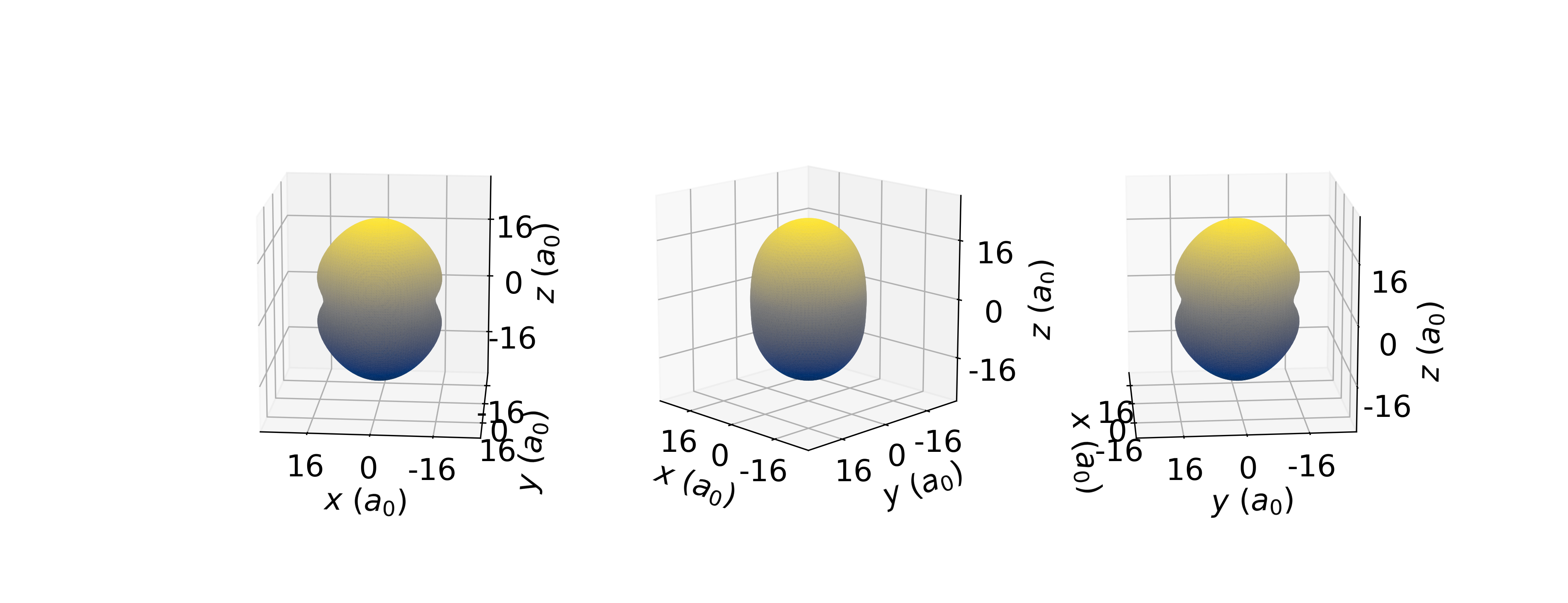}
    \caption{
    4: Polaron density surface $\rho(\mathbf{r}) = 10^{-8}$,
    corresponding to the marker 4 in  Fig.~\ref{fig:supp-phases}.
    $D_{4h}$ point group, $a=1$, $b=0.4$, $c=-0.35$}
    \label{fig:den4}
\end{figure}

\begin{figure}
    \centering
    \includegraphics[scale=0.5]{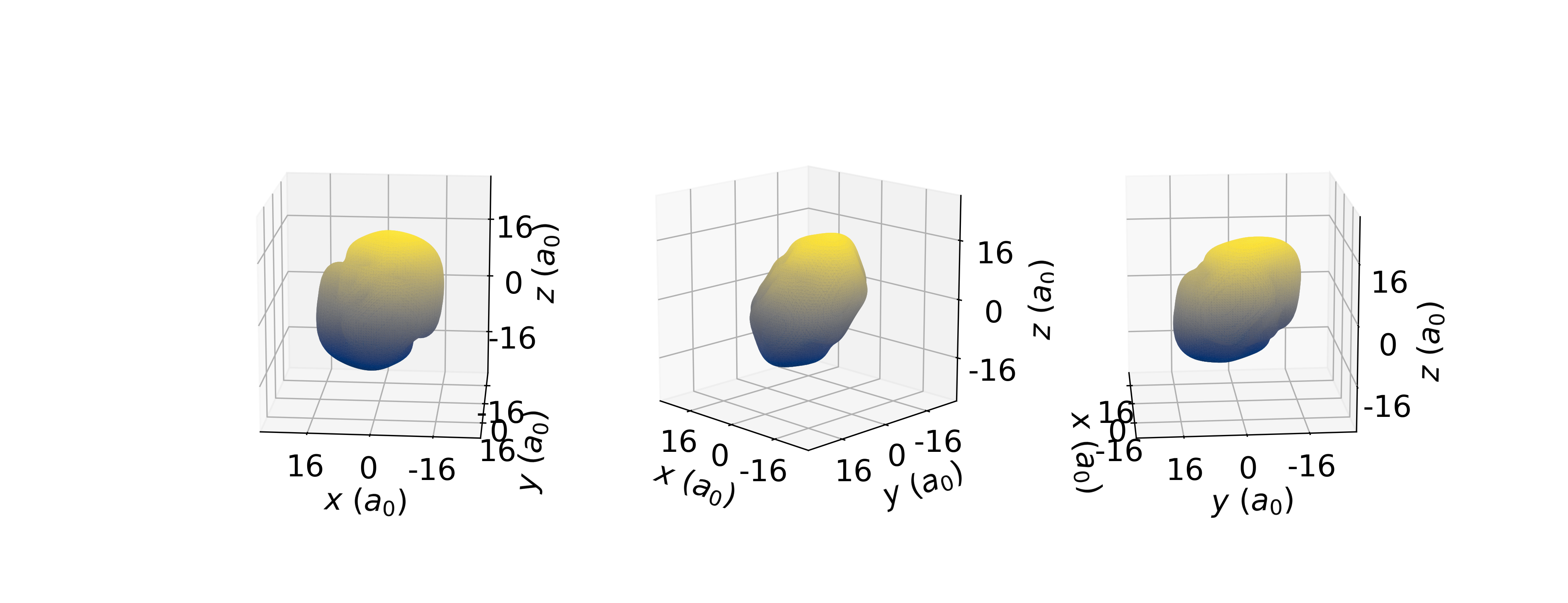}
    \caption{
    5: Polaron density surface $\rho(\mathbf{r}) = 10^{-8}$,
    corresponding to the marker 5 in  Fig.~\ref{fig:supp-phases}.
    $D_{3d}$ point group, $a=1$, $b=0.4$, $c=-0.7$}
    \label{fig:den5}
\end{figure}

\begin{figure}
    \centering
    \includegraphics[scale=0.5]{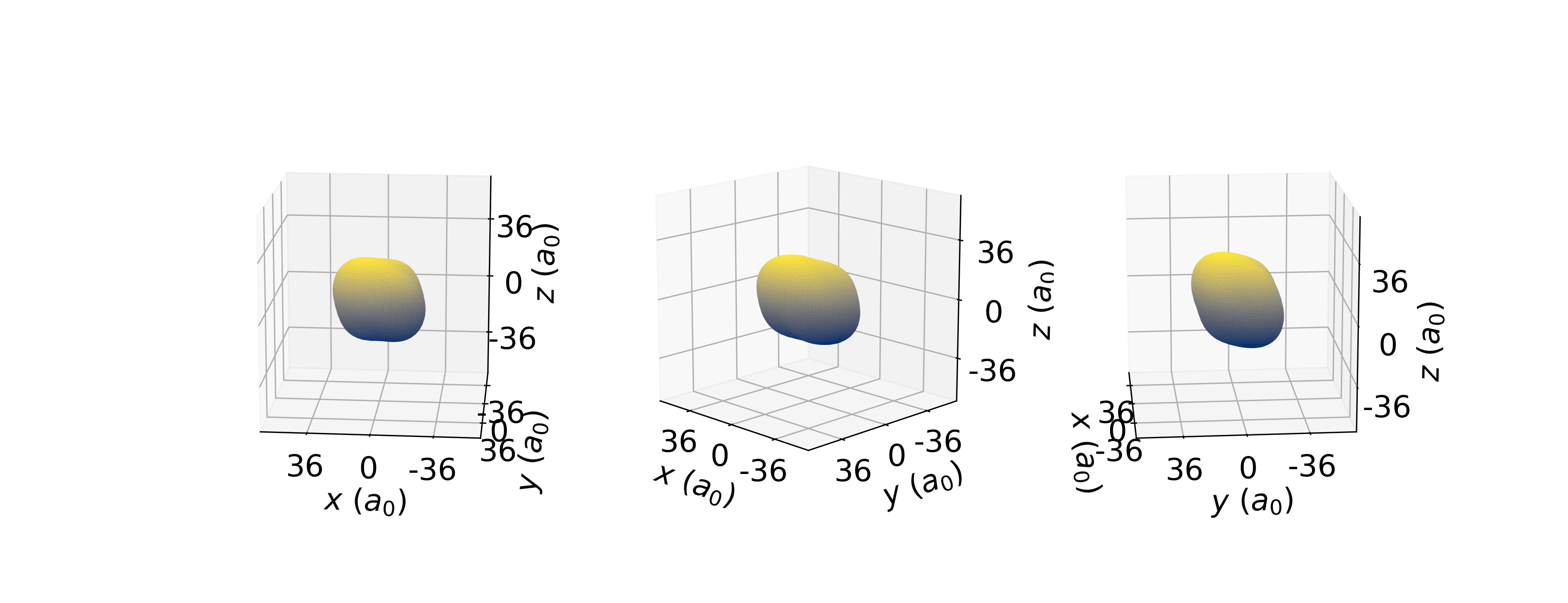}
    \caption{
    6: Polaron density surface $\rho(\mathbf{r}) = 10^{-8}$,
    corresponding to the marker 6 in  Fig.~\ref{fig:supp-phases}.
    $D_{3d}$ point group, $a=1$, $b=1$, $c=1$}
    \label{fig:den6}
\end{figure}

\begin{figure}
    \centering
    \includegraphics[scale=0.5]{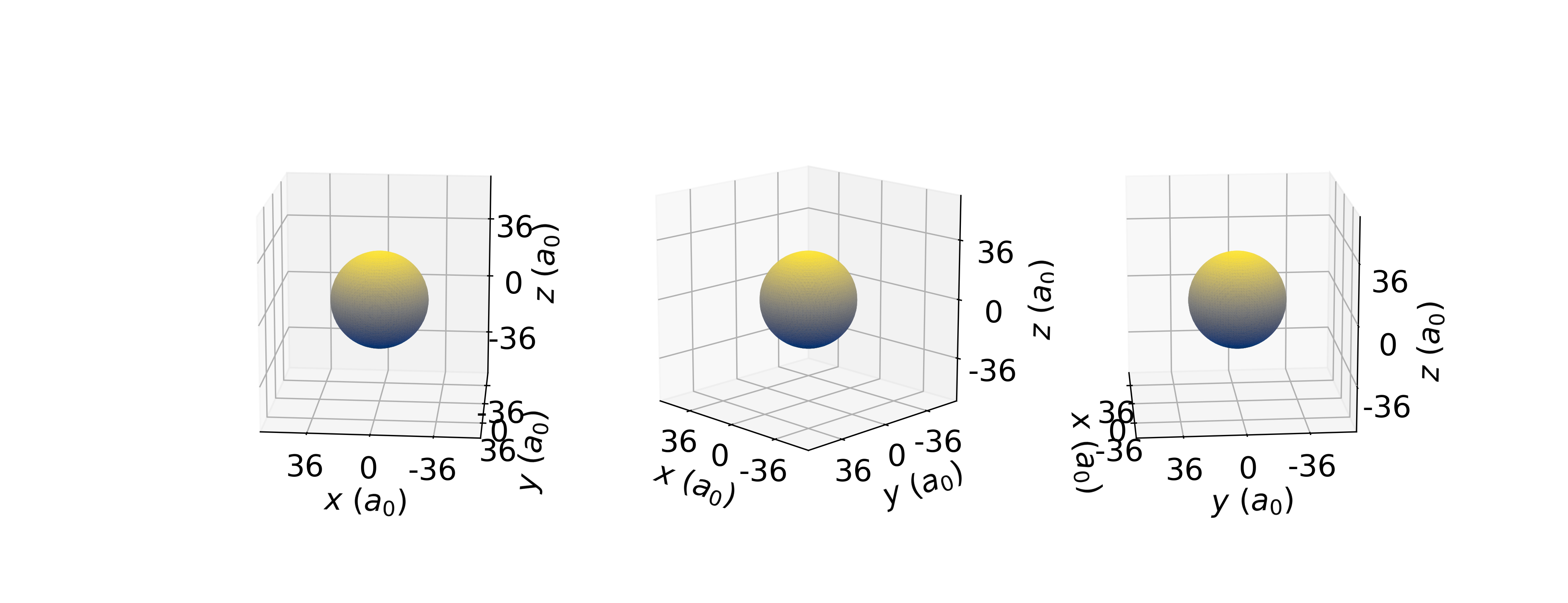}
    \caption{
    7: Polaron density surface $\rho(\mathbf{r}) = 10^{-8}$,
    corresponding to the marker 7 in  Fig.~\ref{fig:supp-phases}.
    $O(3)$ point group, $a=1$, $b=1$, $c=0$}
    \label{fig:den7}
\end{figure}

\begin{figure}
    \centering
    \includegraphics[scale=0.5]{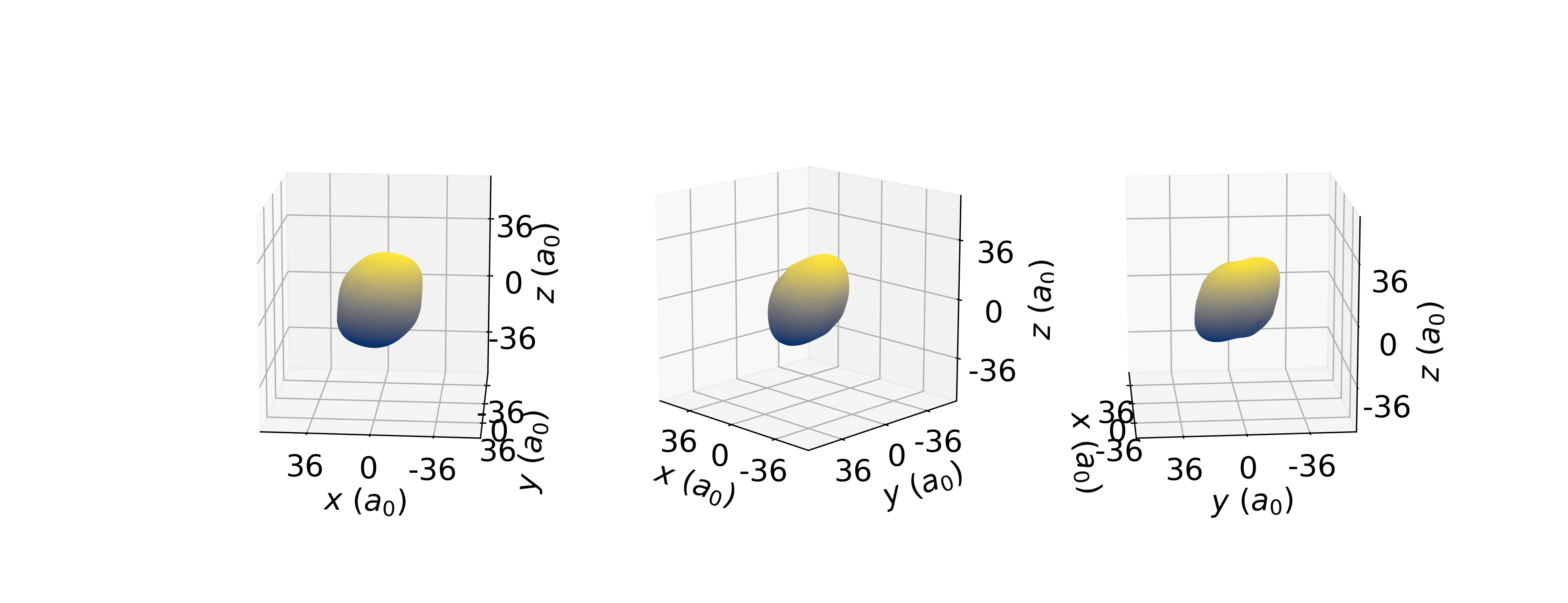}
    \caption{
    8: Polaron density surface $\rho(\mathbf{r}) = 10^{-8}$,
    corresponding to the marker 8 in  Fig.~\ref{fig:supp-phases}.
    $D_{3d}$ point group, $a=1$, $b=1$, $c=-1$}
    \label{fig:den8}
\end{figure}

\begin{figure}
    \centering
    \includegraphics[scale=0.5]{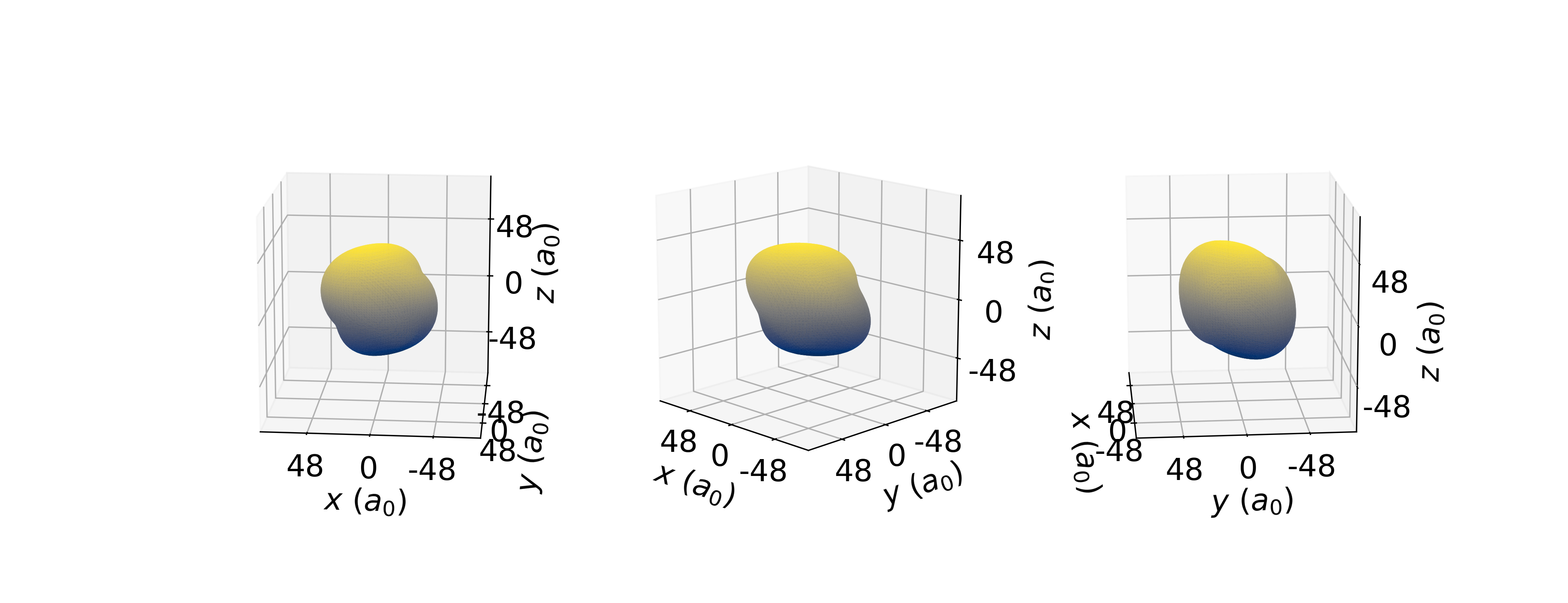}
    \caption{
    9: Polaron density surface $\rho(\mathbf{r}) = 10^{-8}$,
    corresponding to the marker 9 in  Fig.~\ref{fig:supp-phases}.
    $D_{3d}$ point group, $a=1$, $b=2.5$, $c=2.0$}
    \label{fig:den9}
\end{figure}

\begin{figure}
    \centering
    \includegraphics[scale=0.5]{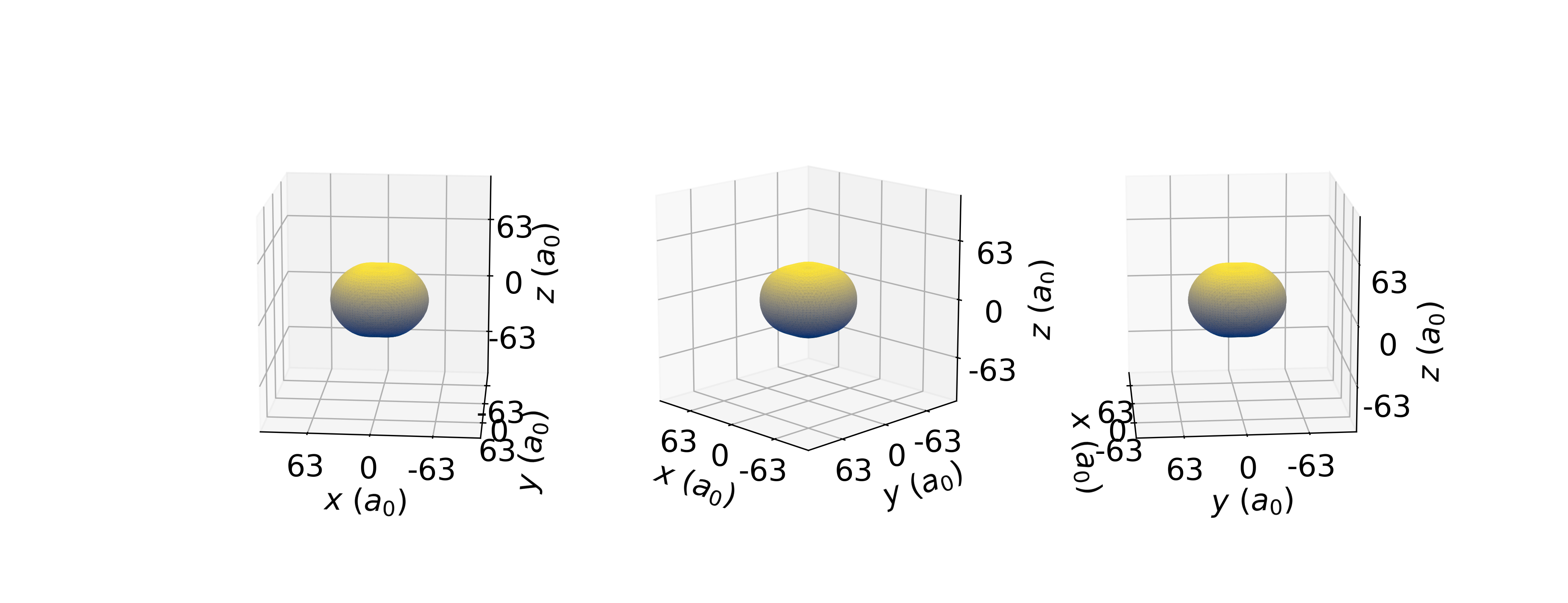}
    \caption{
    10: Polaron density surface $\rho(\mathbf{r}) = 10^{-8}$,
    corresponding to the marker 10 in  Fig.~\ref{fig:supp-phases}.
    $D_{4h}$ point group, $a=1$, $b=2.5$, $c=1$}
    \label{fig:den10}
\end{figure}

\begin{figure}
    \centering
    \includegraphics[scale=0.5]{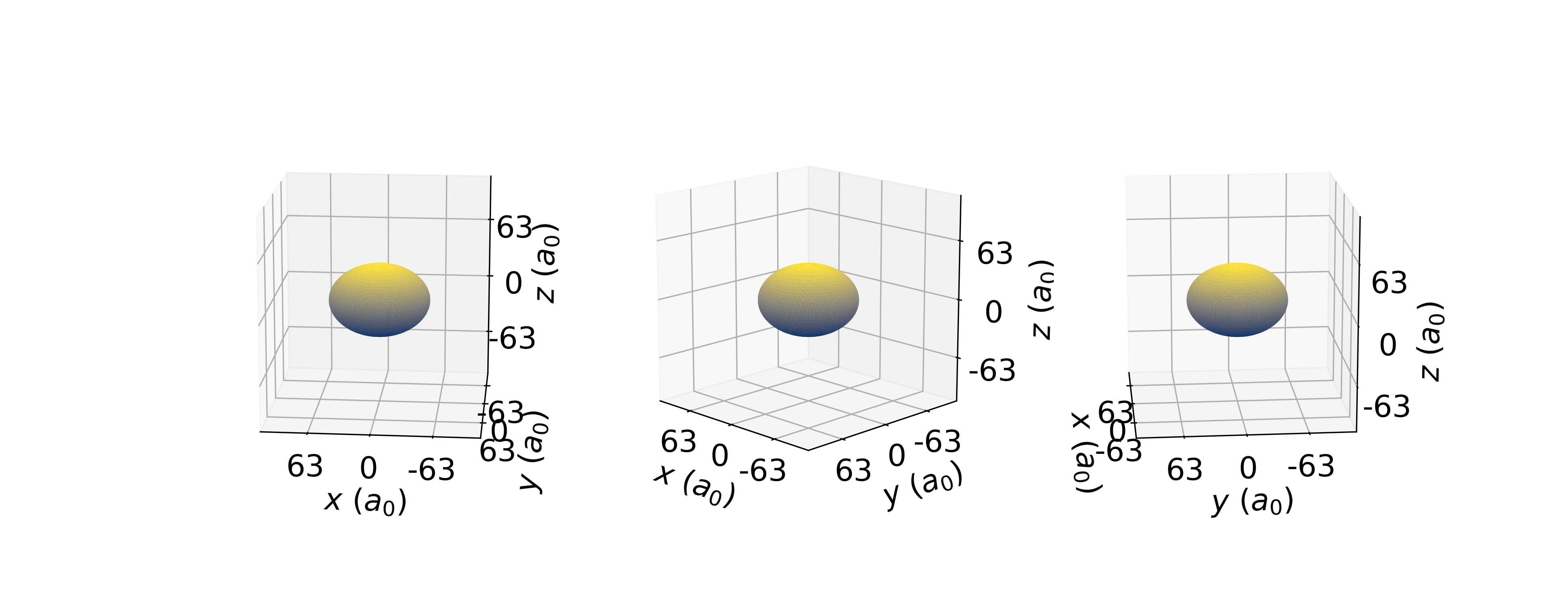}
    \caption{
    11: Polaron density surface $\:rho(\mathbf{r}) = 10^{-8}$,
    corresponding to the marker 11 in  Fig.~\ref{fig:supp-phases}.
    $O(2)$$\times$$O(1)$ point group, $a=1$, $b=2.5$, $c=0$}
    \label{fig:den11}
\end{figure}

\begin{figure}
    \centering
    \includegraphics[scale=0.5]{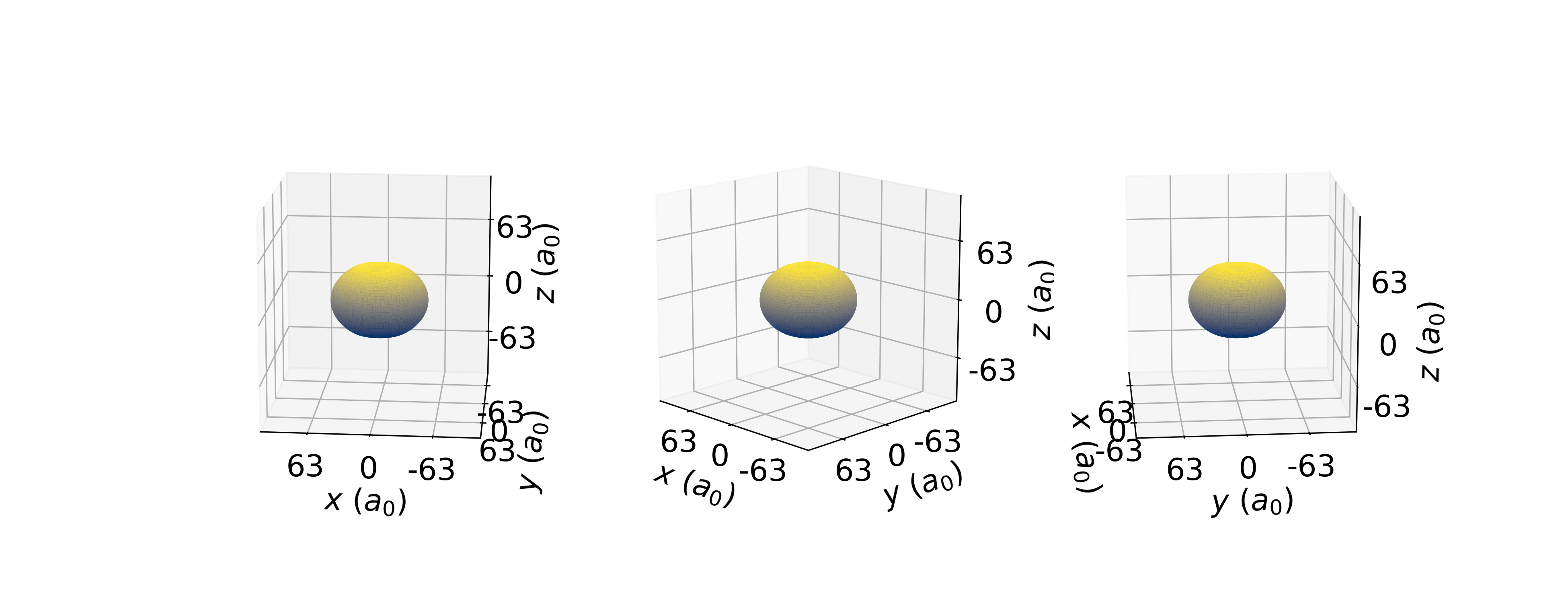}
    \caption{
    12: Polaron density surface $\rho(\mathbf{r}) = 10^{-8}$,
    corresponding to the marker 12 in  Fig.~\ref{fig:supp-phases}.
    $D_{4h}$ point group, $a=1$, $b=2.5$, $c=-1$}
    \label{fig:den12}
\end{figure}

\begin{figure}
    \centering
    \includegraphics[scale=0.5]{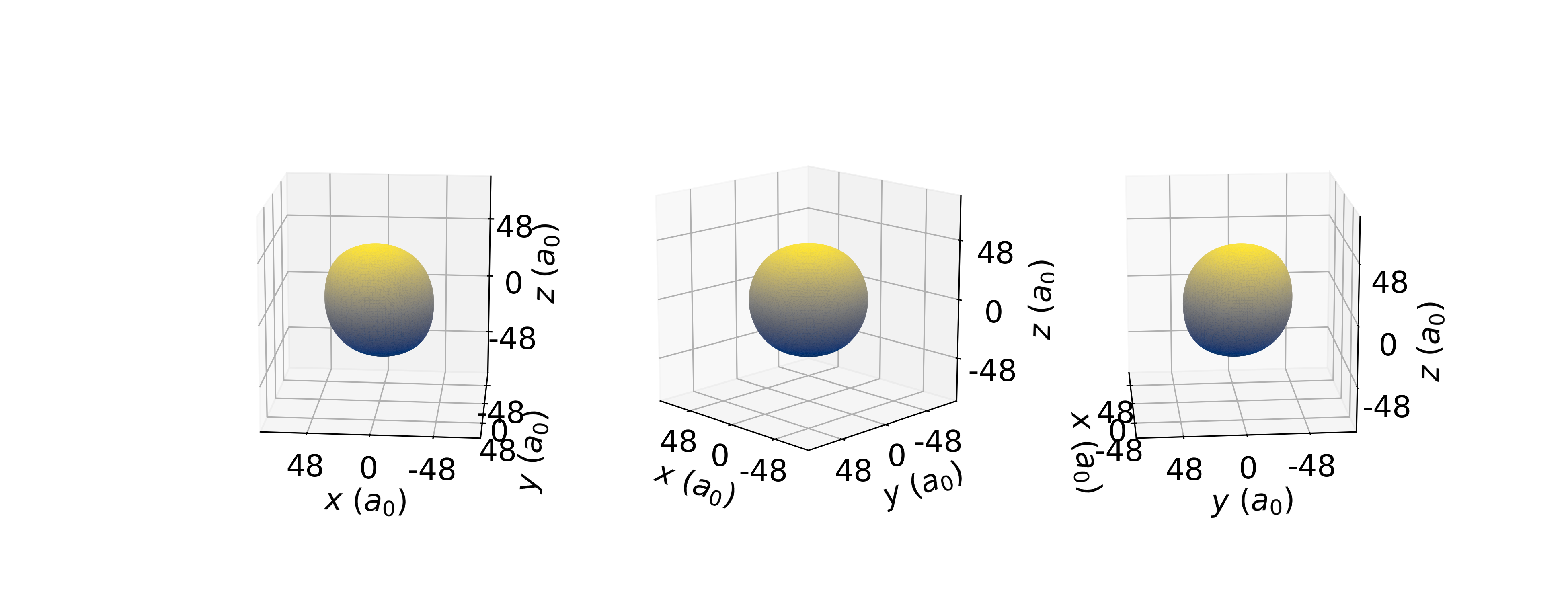}
    \caption{
    13: Polaron density surface $\rho(\mathbf{r}) = 10^{-8}$,
    corresponding to the marker 13 in  Fig.~\ref{fig:supp-phases}.
    $D_{3d}$ point group, $a=1$, $b=2.5$, $c=-2$}
    \label{fig:den13}
\end{figure}

\section{Supplementary tables}
\begin{table}[htbp]
\caption{Parameters of the generalized Fr\"ohlich model for cubic materials with triply degenerate valence band at $\Gamma$-point.
Unit cell length $a_0$, Luttinger-Kohn parameters $a$, $b$, and $c$, high-frequency $\epsilon^\infty$ and static $\epsilon^0$ dielectric constants are taken from Ref.~33.}
\label{table:bogdan2021}
\begin{tabular}{rrrrrrr}
\hline \hline
Material & $a_0$ (Bohr) & $a$ & $b$ & $c$ & $\epsilon^\infty$ & $\epsilon^0$ \\
\hline
AlAs     & 10.825   & -4.681   & -1.019  & -5.498   & 9.49              & 11.51        \\
AlP      & 10.406   & -2.598   & -0.894  & -3.335   & 8.12              & 10.32        \\
AlSb     & 11.762   & -6.473   & -1.372  & -7.520   & 12.02             & 13.35        \\
BAs      & 9.088    & -2.337   & -2.104  & -3.912   & 9.81              & 9.89         \\
BN       & 6.746    & -0.917   & -0.969  & -1.635   & 4.52              & 6.69         \\
CdS      & 11.202   & -3.999   & -0.605  & -4.321   & 6.21              & 10.24        \\
CdSe     & 11.711   & -9.504   & -0.684  & -9.881   & 7.83              & 11.78        \\
CdTe     & 12.513   & -9.517   & -0.867  & -10.033  & 8.89              & 12.37        \\
GaAs     & 10.863   & -54.896  & -1.362  & -55.859  & 15.31             & 17.55        \\
GaN      & 8.598    & -3.392   & -0.555  & -3.762   & 6.13              & 16.30        \\
GaP      & 10.294   & -4.565   & -1.313  & -5.514   & 10.50             & 11.00        \\
SiC      & 8.277    & -1.388   & -0.844  & -2.160   & 6.97              & 10.30        \\
ZnS      & 10.286   & -2.751   & -0.694  & -3.170   & 5.97              & 9.40         \\
ZnSe     & 10.833   & -5.340   & -0.791  & -5.834   & 7.35              & 10.73        \\
ZnTe     & 11.682   & -6.495   & -1.032  & -7.174   & 9.05              & 11.99        \\
CaO      & 9.121    & -1.407   & -0.174  & -0.653   & 3.77              & 16.76        \\
Li$_2$O     & 8.730    & -0.540   & -0.278  & -0.747   & 2.90              & 7.80         \\
MgO      & 8.037    & -1.291   & -0.231  & -1.361   & 3.23              & 11.14        \\
SrO      & 9.810    & -1.519   & -0.120  & -0.625   & 3.77              & 20.91 \\
\hline \hline 
\end{tabular}
\end{table}
\begin{table}[htbp]
\caption{Hole polaron formation energy in the cubic generalized Fr\"ohlich model for a set of real cubic materials, considered in the work.
$E^\mathrm{var}_\mathrm{pol}$ represents formation energies calculated within the fully variational formalism of this work.
$E^\mathrm{gau}_\mathrm{pol}$ denotes formation energies obtained by the Gaussian Ansatz approach of Ref.~33.}
\label{table:bogdan2021_2}
\begin{tabular}{r|rrrr|r}
\hline \hline
Material  & \multicolumn{3}{c}{$E^\mathrm{gau}_\mathrm{pol}$~(meV)} & &
$E^\mathrm{var}_\mathrm{pol}$~(meV) \\
          &   (100)     &    (110)     &    (111)     &              \\
\hline 
AlAs      &   -0.13517  &    -0.20604  &    -0.19736  & &    -0.32545  \\
AlP       &   -0.37606  &    -0.57906  &    -0.54229  & &   -0.81491  \\
AlSb      &   -0.01994  &    -0.02958  &    -0.02875  & &   -0.04661  \\
BAs       &   -0.00023  &    -0.00031  &    -0.00029  & &   -0.00038  \\
BN        &   -3.90519  &    -5.22358  &    -4.90517  & &   -6.32232  \\
CdS       &   -2.27488  &    -3.01849  &    -3.07203  & &   -5.02771  \\
CdSe      &   -0.64406  &    -0.85738  &    -0.87593  & &   -1.92308  \\
CdTe      &   -0.31171  &    -0.42327  &    -0.43000  & &   -0.78815  \\
GaAs      &   -0.00694  &    -0.00960  &    -0.00966  & &   -0.03852  \\
GaN       &   -3.33710  &    -4.69221  &    -4.68922  & &  -15.58637  \\
GaP       &   -0.00647  &    -0.00931  &    -0.00909  & &   -0.01335  \\
SiC       &   -1.54667  &    -2.56839  &    -2.20181  & &   -3.37136  \\
ZnS       &   -2.30872  &    -3.13410  &    -3.14241  & &   -4.51635  \\
ZnSe      &   -0.78833  &    -1.08677  &    -1.09486  & &   -1.85846  \\
ZnTe      &   -0.24936  &    -0.34915  &    -0.34961  & &   -0.58686  \\
CaO       &  -75.87667  &   -62.73596  &   -57.15204  & &  -84.62903  \\
Li$_2$O      &  -96.31797  &  -137.12446  &  -131.20188  & & -175.71030  \\
MgO       &  -77.31154  &   -95.18113  &   -97.82699  & & -143.90013  \\
SrO       & -101.31612  &   -75.53183  &   -64.65848  & & -109.61124  \\
\hline \hline 
\end{tabular}
\end{table}

\begin{table}[htbp]
\caption{
Hole polaron formation energy, $E^{\mathrm{var}}_{\mathrm{pol}}$, obtained through the variational formalism, alongside the split-off energy, $\Delta{\mathrm{SOC}}$ (meV), for a set of materials.
This comparison is provided for cases where the absolute value of the ratio
$R = \left|\Delta_{\mathrm{SOC}} / E^{\mathrm{var}}_{\mathrm{pol}}\right|$ 
is less than 1. 
The current approach, based on the LK Hamiltonian, might be used as a starting point for a more refined analysis, in which the SOC would be treated perturbatively.
However, even in this case, the symmetry of the ground-state polaron will likely not be the same as the one determined from the present one. For other materials, the ratio $R$ is much larger than 1.
The SOC coupling dominates over the polaron energy. 
The present approach, based on the LK Hamiltonian is not valid anymore, as the starting electronic structure is far from the real one.}
\label{table:soc}
\begin{tabular}{r|rr|r}
\hline \hline
Material  & $E^\mathrm{var}_\mathrm{pol}$~(meV) & $\Delta_\mathrm{SOC}$~(meV) & $R$\\
\hline 
Li$_2$O & -175.71030 &  33.47002 &  0.191 \\
 MgO & -143.90013 &  37.82384 &  0.263 \\
 CaO &  -84.62903 &  40.54498 &  0.479 \\
 SrO & -109.61124 &  60.13719 &  0.549 \\
 GaN &  -15.58637 &  13.60570 &  0.873 \\
\hline \hline 
\end{tabular}
\end{table}
